\documentclass[aps,prb,twocolumn,superscriptaddress,nofootinbib,longbibliography]{revtex4-2}

\usepackage{amsmath,amssymb}
\usepackage{amsfonts}
\usepackage{slashed}	
\usepackage{soul}
\usepackage{braket}

\usepackage[export]{adjustbox}

\usepackage{graphicx}
\usepackage{fancybox,xcolor}
\usepackage{float}

\usepackage[utf8]{inputenc} 
\usepackage[T1]{fontenc} 

\usepackage[font=footnotesize,labelfont=bf, justification=centerlast]{caption}
\usepackage[normalem]{ulem}
\usepackage{xspace}
\usepackage{enumitem}

\usepackage[english]{babel}
\usepackage{lmodern}
\usepackage{placeins} 

\usepackage{amsthm}
\theoremstyle{remark}

\def\ra{\rightarrow}

\def\hc{\text{h.c.}}

\def\1{\mathbf{1}}
\renewcommand{\Re}{\mbox{Re}}
\renewcommand{\Im}{\mbox{Im}}

\def\Ocal#1{{\mathcal O}\left(#1\right)}

\def\F{{\cal F}}
\def\G{{\cal G}}
\def\L{{\cal L}}
\renewcommand{\ell}{L}
\def\M{{\cal M}}

\def\hc{{\rm{h.c.}}} 
\def\>{\rangle}
\def\<{\langle}
\newcommand{\ev}[1]{\Braket{#1}}
\newcommand{\ketbra}[2]{\Ket{#1}\!\!\Bra{#2}}
\def\p{\partial}

\def\t{\widetilde}

\def\s{{}^\dagger}

\def\eps{\varepsilon}
\def\al{\alpha}

\def\d{\delta}
\def\D{\Delta}

\def\Ga{\Gamma}
\def\su{\uparrow}
\def\sd{\downarrow}
\def\T{T_\text{B}}
\def\kB{k_\text{B}}
\def\Si{\mbox{Si}}

\begin{document}

\title{Effective time--dependent temperature for fermionic master equations\\
beyond the Markov and the secular approximations}

\author{Lukas Litzba}
\email{lukas.litzba@uni-due.de}
\affiliation{Fakult\"at f\"ur Physik and CENIDE, Universit\"at Duisburg-Essen, Lotharstra{\ss}e 1, Duisburg 47057, Germany}
\author{Eric Kleinherbers}
\email{ekleinherbers@physics.ucla.edu}
\affiliation{Department of Physics and Astronomy, University of California, Los Angeles, California 90095, USA}
\affiliation{Fakult\"at f\"ur Physik and CENIDE, Universit\"at Duisburg-Essen, Lotharstra{\ss}e 1, Duisburg 47057, Germany}
\author{J\"urgen K\"onig}
\affiliation{Fakult\"at f\"ur Physik and CENIDE, Universit\"at Duisburg-Essen, Lotharstra{\ss}e 1, Duisburg 47057, Germany}
\author{Ralf Sch\"utzhold}
\affiliation{Helmholtz-Zentrum Dresden-Rossendorf, Bautzner Landstra{\ss}e 400, 01328 Dresden, Germany}
\affiliation{Institut f\"ur Theoretische Physik, Technische Universit\"at Dresden, 01062 Dresden, Germany}
\author{Nikodem Szpak}
\email{nikodem.szpak@uni-due.de}
\affiliation{Fakult\"at f\"ur Physik and CENIDE, Universit\"at Duisburg-Essen, Lotharstra{\ss}e 1, Duisburg 47057, Germany}

\date{\today}

\begin{abstract}
  We consider a fermionic quantum system exchanging particles with an environment at a fixed temperature and study its reduced evolution by means of a Redfield--I equation
  with time--dependent (non–Markovian) coefficients.
  We find that the description can be efficiently reduced to a standard--form Redfield--II equation, however, with a \textit{time--dependent effective bath temperature} obeying a universal law.
  At early times, after the system and environment start in a product state, the effective \textit{temperature} appears to be very high, yet eventually it settles down towards the true environment value.
  In this way, we obtain a time--local master equation, offering high accuracy at all times and preserving the crucial properties of the density matrix.
  It includes non–Markovian relaxation processes beyond the secular approximation and time–averaging methods and can be further applied to various types of Gorini--Kossakowski--Sudarshan--Lindblad equations.
  We derive the theory from first principles and discuss its application using a simple example of a single quantum dot.
\end{abstract}

\maketitle

\section{Introduction} \label{sec:Intro}

When dealing with open quantum systems \cite{Breuer} interacting with an infinite environment, exact solutions can be found in some special cases 
\cite{meir1992a,ExactTimeDepPhysRevB.50.5528,karrlein1997a,PhysRevB.75.195115,Zedler_Schaller_PhysRevB.80.045309,ExactNonMarkovDyPhysRevLett.109.170402,Topp_2015,exactPhotonenPhysRevA.95.033830,JussiauPhysRevB.100.115411,jussiau2020multiple}.
However, as soon as any interactions, such as Coulomb, are present
it becomes almost impossible to solve them exactly \cite{meir1992a,kolovsky2020a,nakagawa2021a,queisser2019a}.
In order to explore the system alone, it is possible to eliminate the baths from the description and obtain a formally exact time--nonlocal (non--Markovian)   
master equation \cite{nakajima1958quantum, zwanzig1960ensemble,koenig_1996,koenig_1996_2,timm_2008,FermionicStocasticME} for the quantum system which {contains information about the full history
including the formation of coherences between the environment and the system} \cite{Zedler_Schaller_PhysRevB.80.045309,Breuer}. 

At the lowest order of perturbation theory in the system--bath coupling, 
using the (first) Markov assumption,
it can be approximated by the time--local master equation for the system's density matrix with time--dependent (non--Markovian) coefficients, known as the Redfield--I equation \cite{Breuer,Redfield1957}.
It does not include the history of the system and of the bath but its time--dependent coefficients maintain residual information about the formation of coherences
between the environment and the system. 
The latter is related to non--Markovian effects present also in unstructured environments (e.g. such as the wideband limit) which are usually treated as Markovian.  
Although the Redfield--I equation offers a good approximation of the system dynamics \cite{hartmann2020a} it is also known for its mathematical problems,
originating in the first order time--dependent perturbation theory,
of not preserving the positivity of the density matrix
which may result in negative probabilities and non--physical behavior of observables
\cite{dumcke1979proper,suarez1992memory,gaspard1999slippage,B07-Dimer, hartmann2020a, LL-MTh}.

Partially these problems are related to 
{the transition rates of the system described by}
the time--dependent coefficients, 
which we denote symbolically $F_{\D E, \T}(t)$ (for transition energy $\D E$ and temperature $\T$),
showing excessive oscillations and becoming temporarily negative (problem \textbf{1}).
In order to remove the oscillations, 
an additional approximation extends the initial integration time to the infinite past ($t_0 \ra -\infty$) by which the resulting Liouville operator becomes time--independent and gives the (now Markovian) Redfield--II master equation 
with static and positive coefficients $F_{\D E, \T}(\infty)$. 
Those, however, show another problem {(the more popular one of the two, regarding literature)}:
{not only single coefficients $F_{\D E, \T}$ but also their 
[matrix--valued] combinations, involved in generic transitions with different transition energies $\D E$, 
can lead to excess coherences between energy states and thus also violate the positivity of the density matrix (problem \textbf{2})
\cite{dumcke1979proper,suarez1992memory,gaspard1999slippage,B07-Dimer, hartmann2020a, LL-MTh}.
}

A standard, however drastic, way to deal with the latter problem is the secular approximation \cite{Breuer, Davies1} that  
{artificially removes energy coherences from the system by which the master equation attains}
the Gorini--Kossakowski--Sudarshan--Lindblad (GKSL) form which preserves all important properties of the density matrix  
\cite{Breuer,Lindblad1976, gorini1976completely,GKLSHistory}. 
However, the secular approximation is known to miss some important physical information about coherences {between energy states} in the system, as demonstrated e.g. in \cite{Global-Local-Sec-2HarmOsc, B07-Dimer,SynchronizedEric,LL-MTh,B07-Trimer}.
Along similar lines as in \cite{Wichterich_coh,Kirsanskas+Wacker-CohLindPhenom, Davidovic2020, Nathan-CohLindDerivation}, in \cite{B07-Dimer, B07-Trimer} we 
proposed 
a refined method of \textit{coherent approximation} which allows to keep the mathematically maximal amount of coherences {in the system}, also leading to a GKSL equation. 
In addition, there exist further regularization methods \cite{Davidovic2020,Potts_2021,d2023time,BeckerPhysRevE.104.014110,TrushechkinPhysRevA.103.062226} which also lead to GKSL master equations.
Due to their simplicity and direct interpretation of the jump operators, the GKSL master equations can be usually derived from phenomenological \cite{Kirsanskas+Wacker-CohLindPhenom, B07-Dimer} or microscopic \cite{hartmann2020a, B07-Dimer,Davidovic2020, Potts_2021,d2023time} points of view.

However, due to their Markovian behavior, {the GKSL equations} neglect memory effects and relaxation dynamics of the environment as well as effects related to the formation of coherences between the environment and the system.
In particular, neglecting the time dependence of the coefficients leads to a loss of accuracy at short times
which can be partly compensated by the ``initial slip'' method, {artificially adjusting the initial state of the system to the later development} 
\cite{suarez1992memory,gaspard1999slippage,Whitney-StayingPositive}. 
The behavior at short times has been also discussed in the context of time–local master equations 
 \cite{Davidovic2020,d2023time,SchallerCoarseGraining,NMdynamicsPhysRevResearch.4.043075,PhysRevX.11.021041}. 
One popular approach which deals with the problem of time--dependent coefficients 
is the dynamical coarse graining (DCG) method \cite{SchallerCoarseGraining}.
It improves the behavior at short times and preserves the properties of the density matrix. 
However, due to the averaging character of the coarse graining, this method becomes similar to the secular approximation for late times, 
affected by the above mentioned problems.

In {Ref.~}\cite{hartmann2020a}, various  master equations have been compared with exact solutions. 
The general conclusion was that ``the simple Redfield--I equation with time--dependent coefficients is significantly more accurate than all other methods''.
Therefore, we take it as a natural starting point {to study a generic tunnel coupling between the system and a fermionic environment}.
Studying the time--dependent coefficients $F_{\D E, \T}(t)$ in more detail, we realized that their defining integrals, parameterized by the time $t$, temperature $\T$, and energy difference $\D E$, can be very accurately and uniformly in all three parameters approximated by a simple family of functions $\F_{\D E, \T}(t)$, obtained by approximate calculation of their defining integral. 
A surprising observation is that the result has again the form of the static coefficients
{$F_{\D E, T(t)}(\infty)$ with now time--dependent temperature 
\begin{equation} 
  T(t) = \T \Bigl/\tanh\left(\frac{4\kB \T}{\hbar \pi}(t-t_0)\right).
\end{equation}
It is universal in the sense that it} depends only on the true bath temperature $\T$, the Boltzmann and Planck constants $\kB$ and $\hbar$, the initial time $t_0$ and time $t$
but not on the energy differences $\D E$ nor on any details of the system or the coupling.
It has the properties that $T(t_0) \ra \infty$ and $T(\infty) \ra \T$. 
The replacement $T_\text{B} \ra T(t)$ leads to a modified Redfield--I equation with time--dependent coefficients, known analytically for all values of parameters $\D E, \T$ and $t$ and satisfying all relevant limiting cases.
{Most importantly, the time-dependent temperature solves problem \textbf{1} since it removes both excessive oscillations and negative values by ensuring $0\le F_{\D E, T(t)}(\infty)\le 1$. It can be naturally combined with the regularisation methods, discussed above, to solve problem \textbf{2} thus bringing the equation into a time--dependent GKSL form.}

The time--dependent temperature $T(t)$ can be interpreted as an effective bath temperature from the perspective of the system. It is conceptually different from time--dependent quasi--equilibrium temperatures of the system discussed in \cite{hebel1959nuclear,Primakoff_PhysRev.130.1267,andersen1964exact, T(t)systemPhysRevA.64.052110(2001)} or time--dependent bath temperatures for finite--size baths discussed in \cite{Schaller_2014,ME_mesoReserviorPhysRevE.105.054119(2022),mesoReserviorPhysRevLett.131.220405(2023)}. 
In our case, the time--dependence of $T(t)$ is related to the energy gain ($T(t)\geq \T$) resulting from the coupling between the system and the bath.

The main goal of this paper is to propose a universal approach in the form of a time--local master equation, offering high accuracy at all times and 
preserving the properties of the density matrix,
which includes non--Markovian relaxation processes 
beyond the secular approximation and time--averaging methods. 

\section{The model}

In the following, we will focus on fermionic systems and fermionic environments
which can exchange particles. 
We consider a finite quantum system described by the Hamiltonian written in its Fock--eigenbasis
\begin{equation}
    H_\text{S}=\sum_{l=1}^{\dim(S)} E_l \ket{E_l}\bra{E_l}
    \label{HSeigen}
\end{equation}
coupled to $M$ fermionic baths, described by the Hamiltonian
\begin{equation}
    H_\text{B} = \sum_{m=1}^M\sum_k \eps_{m,k}\, b\s_{m,k}\, b_{m,k},
    \label{HBintro}
\end{equation}
via the coupling Hamiltonian 
\begin{equation}
    H_\text{C} = \sum_{m=1}^M\sum_k \gamma_{m,k}\, c\s_{m} \otimes\, b_{m,k} + \hc
    \label{HCintro}
\end{equation}
{(brought to the tensor product form via the Jordan--Wigner transformation \cite{schaller2014open})}.
Here,
$c\s_{m}$ and $c_{m}$ are fermionic creation and annihilation operators in the system%
\footnote{The case where more than one bath, e.g. $m=1,2,...,M_1\leq M$, is coupled to the same system mode can be described by the formal substitution $c_1=c_2=...=c_{M_1}$.}, respectively, 
whereas $b\s_{m,k}$ and $b_{m,k}$ are the fermionic creation and annihilation operators, respectively, at bath $m$ in the mode $k$ which are associated with the energy $\eps_{m,k}$ and satisfy $\{b_{m,k}, b\s_{n,l}\} = \d_{mn} \d_{kl}$, $\{b_{m,k}, b_{n,l}\} = 0$. 
The coefficients $\gamma_{m,k}$ give the {tunneling amplitudes} between the system and the mode $k$ in the bath $m$. 
The total Hamiltonian reads then
\begin{equation}
   H = H_S \otimes \mathrm{1}_\text{B} + \mathrm{1}_\text{S} \otimes H_\text{B} + H_\text{C}.
   \label{Hgrund}
\end{equation}
We assume that the dimension of the system,
$\dim(S)$,
is small compared to the number of degrees of freedom of the baths, $\dim(S)\ll \dim(B)$.
This justifies the treatment of the system as open, coupled to the much larger environment, which shall be subsequently eliminated from the description. 
We assume also that the full system begins its evolution at time $t=t_0$ in a product state described by the density matrix
$ \rho(t_0) = \rho_\text{S}(t_0) \otimes \rho_B(t_0) $ where $\rho_S(t_0)$ refers to the system while 
\begin{equation}
    \rho_B(t_0) \sim e^{-\left(H_\text{B}-\mu  \sum_{m,k}b^\dagger_{m,k}b_{m,k}\right)/(\kB \T)}
\end{equation}
refers to the baths at thermal equilibrium (Gibbs state) with temperature $\T$ and chemical potential $\mu$ and satisfies $[H_B, \rho_B] = 0$. 
For unequal chemical potentials $\mu_m$ and temperatures $T_{\text{B},m}$, the presented method can be applied analogously giving rise to separate effective temperatures $T_m(t)$ for each bath $m$.

For a Hamilton operator of the form (\ref{Hgrund}), in section \ref{sec:TimeDepRed}, we will derive a time--dependent Redfield equation including time--dependent coefficients. 
In section \ref{sec:TimeDepCoef}, we will discuss these coefficients and identify that their excessive oscillations lead to non--physical effects. 
Furthermore, we will demonstrate that the time--dependent coefficients can be interpreted, to a very good approximation, as the static coefficients with a \textit{time--dependent effective} temperature, $T(t)$.
Using this interpretation, we will find a time--local master equation with positive time--dependent coefficients.

As the simplest example, demonstrating the application of the method, 
we will consider a single quantum dot with Coulomb interaction coupled to a fermionic bath in which the problems of the time--dependent Redfield--I equation become already apparent. 
In section \ref{sec:singleDot}, we will compare the solutions of the Redfield--I equation to its version using the time--dependent temperature and to static GKSL equations. 
In section \ref{sec:benchmark},
we will compare the above approximation schemes with exact solutions obtained for a non--Coulomb--interacting quantum dot.

\section{Time--dependent Redfield equation}\label{sec:TimeDepRed}

In order to effectively eliminate the environment from the description, we apply the Born and the first Markov approximations to the bath and the system evolution and, by tracing out the baths' degrees of freedom in the von Neumann equation expanded to the lowest non--vanishing order in the system--bath couplings, $\Ocal{|\gamma_{m,k}|^2}$, 
we arrive at the Redfield--I master equation \cite{Breuer,Redfield1957}
\begin{multline} \label{vonNeumann}
    \p_t\rho_\text{I}(t)=\\-\int_{0}^{t-t_0}\text{d}\tau \text{ } \text{tr}_\text{B}[H_\text{C,I}(t),[H_\text{C,I}(t-\tau),\rho_\text{S,I}(t)\otimes\rho_\text{B}(t_0)]].
\end{multline}
The index ``I'' indicates the interaction picture with respect to the coupling Hamiltonian $H_C$. Furthermore, we set $\hbar = 1$ for convenience.

With the baths in thermal equilibrium, satisfying  $\ev{b^\dagger_{m,k}b_{m',k'}} =\delta_{k,k'}\delta_{m,m'}f_+(\eps_{m,k}, \T)$, and the Fermi function
 $f_\pm(E, \T) = [1+\exp\{\pm(E-\mu)/(\kB \T)\}]^{-1}$, the Redfield--I master equation, now in the Schrödinger picture, can be also written in the form
\begin{equation} \label{Redfied-I}
   \p_t\, \rho(t) = -i\, [H_S, \rho(t)] + \L_t\, \rho(t)
\end{equation}
with the superoperator
$\L_t$ {acting in the full Liouville space} 
which is, in general, not positivity preserving%
\footnote{Positivity of the superoperator $\L$ means the map $\mathcal{U}_t \equiv \hat{T} \exp\big(\!\int_0^t\text{d}\tau\L_\tau\! \big)$ preserves the positivity of the density matrix $\rho(t) = \mathcal{U}_t \rho(0)$ for all $t \geq 0$. It is also required that $\L$ preserves the trace of $\rho$.}.
It can be split into two parts, $\L_t = -i [\d H_S(t), \cdot] + \t \L_t$, of which the first can be included in the ``renormalized'' (or ``Lamb-shifted'') hermitian, possibly time--dependent Hamiltonian 
\begin{equation} \label{H-ren}
\begin{split}
	&\t H_S(t) = H_S + \d H_S(t) = H_S \\
	&- i \sum_{\substack{m,\al \\ \D E, \D E'}} \frac{A^{\alpha}_{m}(t,\D E) - \overline{A^\alpha_m(t,\D E')}}{4} 
	K^\al_{m}(\D E')^\dagger K^\al_{m}(\D E)
\end{split}
\end{equation}
with 
\begin{equation}
   K^\pm_{m}(\D E) = \sum_{k,l} \d_{E_k-E_l,\pm \D E} \ketbra{E_k}{E_k} c^\pm_{m} \ketbra{E_l}{E_l}
\end{equation}
and 
\begin{equation} \label{F(t)}
\begin{split}
      &A^\pm_m(t,\D E) = \\
      &= 2\int_0^{t-t_0}\text{d}\tau\sum_{k} |\gamma_{m,k}|^2 f_\pm(\eps_{m,k}, \T)\, e^{\pm i (\eps_{m,k} - \D E) \tau} \\
      &= \int_0^{t-t_0}\text{d}\tau\int_{-\infty}^\infty\frac{\text{d}\omega}{\pi} \Ga_{m}(\omega) f_\pm(\omega, \T)\, e^{\pm i (\omega - \D E) \tau}
\end{split}
\end{equation}
where we introduced $c^-_m=c_m$, $c_m^+=c_m^\dagger$ and $\Ga_{m}(\omega) = 2 \pi \sum_k |\gamma_{m,k}|^2 \d(\omega-\eps_{m,k})$. Since the bath's spectrum should be dense, 
{in the following we will assume that $\Ga_{m}(\omega)$ becomes a continuous function with an effective bandwidth $\D_\Ga$ \cite{LL-MTh}. 
Since this will restrict our further considerations to times $t-t_0\gg 1/\D_\Ga$ 
it should be assumed that $1/\D_\Ga$ is shorter than any other relevant timescale.
For simplicity of the presentation, we will consider here only the wideband limit  $\D_\Ga \ra \infty$ with constant $\Ga_{m}(\omega)=\Gamma_{m}$ for all energies $\omega$
(cf. App.~\ref{app:nonwideband} for a discussion). 
}
$\D E$ 
refers to all possible differences of eigenenergies of the system Hamiltonian $H_S$ while 
$\al = \pm$ refers to creation ($+$) and annihilation ($-$) processes. 

The Redfield--I equation becomes
\begin{equation} \label{Redfield-mod}
	\p_t \rho = -i [\t H_S, \rho] + \t \L_t \, \rho
\end{equation} 
with the new Liouville superoperator
\begin{equation} \label{LL}
	\t \L_t \rho = 
	\sum_{\substack{m,\al \\ \D E, \D E'}} M^\al_{m}(t, \D E, \D E')\, \L_{m}^\al(\D E, \D E')\, \rho,
\end{equation}
including the coefficients
\begin{equation} \label{M=F+F}
	M^\al_{m}(t, \D E, \D E') = \frac{A^{\alpha}_{m}(t,\D E) + \overline{A^\alpha_m(t,\D E')}}{2}
\end{equation}
and the superoperators
\begin{multline}
	\L_{m}^\al(\D E, \D E') \rho= 
	K^\al_{m}(\D E)\, \rho\, K^\al_{m}(\D E')^\dagger \\
	 - \frac{1}{2} \left\{ K^\al_{m}(\D E')^\dagger K^\al_{m}(\D E), \rho \right\}.
\end{multline}
The indices $m, \alpha$ refer to the effective relaxation channels.
The total Liouville superoperator $\t \L_t$ does not necessarily preserve the positivity of $\rho$, as explained in Sec.~\ref{sec:Intro}. 
The coefficients $M^\al_{m}(t, \D E, \D E')$ are time--dependent due to integration 
over a finite history 
between the starting point at $t=t_0$, when the system and the bath were prepared in a product state, and the current time $t$. 

In the standard scheme, the coefficients are stabilized 
by the (second) Markov approximation, shifting the starting point of the evolution to $t_0 \rightarrow -\infty$.
By integrating over the infinite history, the theory becomes Markovian.
The coefficients reduce to constants depending only on the 
Fermi distribution of the bath
leading to the Redfield--II equation.
It provides the staring point for further approximations, e.g. the secular, coherent or other approximations \cite{Kirsanskas+Wacker-CohLindPhenom, Davidovic2020, Nathan-CohLindDerivation, B07-Dimer, B07-Trimer,Davidovic2020,Potts_2021,d2023time,BeckerPhysRevE.104.014110,TrushechkinPhysRevA.103.062226}, leading to various versions of the GKSL equation (cf. App.~\ref{app:RedToLind}). 

\section{Time--dependent temperature}\label{sec:TimeDepCoef}

\subsection{Real part of $A^\pm_m(t,\D E)$}\label{sec:TimeDepCoefRE}
\begin{figure}[t]
    \includegraphics[width=\linewidth]{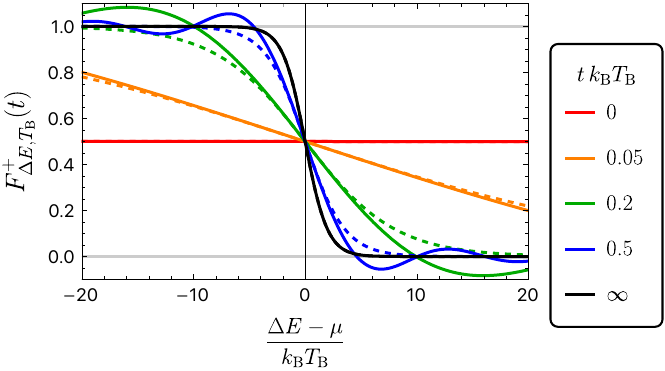}
    \includegraphics[width=\linewidth]{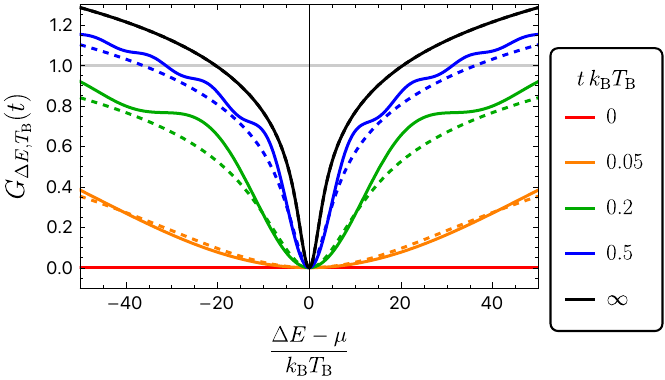}
  \caption{Solid lines show the real part $F^+_{\D E, \T}(t)$ (top) and the imaginary part $G_{\D E, \T}(t)$ (bottom) of the coefficients {$A^+_m(t,\D E)$} as functions of $\Delta E$ in units of $\kB T$.
  The dashed lines show the approximations based on the time-dependent temperature $T(t)$, the real part $\F^+_{\D E, \T}(t)$ (top) and the imaginary part ${\G}_{\Delta E, \T}(t)$ (bottom). 
  Different colors correspond to various times $t$.
  For $t=0$ (red) and $t\rightarrow\infty$ (black), the solid and dashed lines match together.
  The original real parts $F^+_{\D E, \T}(t)$ overshoot the interval $[0,1]$ while their approximations stay within it.}
  \label{fig:F(DE)}
\end{figure}

Here, we first stay with the Redfield--I equation and look closer at the time--dependence of the coefficients $A^\pm_m(t, \D E)$.
In the wideband limit {(for non--wideband cf. App. \ref{app:nonwideband})}, the real part of (\ref{F(t)}) becomes 
\begin{equation}  \label{ReA}
    \Ga_m F^\pm_{\D E, \T}(t)=\Re\,A^\pm_m(t,\D E) 
\end{equation}
where
\begin{multline} \label{ReF}
    F^\pm_{\D E, \T}(t) \equiv 
    \frac{1}{2} \mp \kB \T\int_{0}^{t-t_0} \frac{\sin[(\D E - \mu)\tau]}{ \sinh(\pi \kB \T\tau)} d\tau.
\end{multline}
For $\kB \T \lesssim |\D E-\mu|/4$ the factors $F^\pm_{\D E, \T}(t)$ show excessive 
oscillations and become negative or larger than one for times $t \sim 1/ |\D E-\mu|$ (cf. solid lines in Fig.~\ref{fig:F(DE)}, top). 
This leads to the above mentioned problems of the Redfield equation
(cf. Sec.~\ref{sec:Intro}) resulting in violation of the positivity of the density matrix (cf. also Sec.~\ref{sec:singleDot} for particular examples). 

At the initial time, 
$F^\pm_{\D E, \T}(t=t_0) = \frac{1}{2}$ is independent of the energy difference $\D E$ and temperature $\T$. 
This corresponds to the Fermi function $f_\pm(\D E, {\infty})~=~\frac{1}{2}$ at infinite temperature. 
The Redfield equation (\ref{Redfield-mod}) simplifies then to the GKSL equation (cf. App. \ref{app:RedToLind}) coupled to {infinitely} hot baths
with 
$L_{m,-}=c_{m}/\sqrt{2}$ and $L_{m,+}=c_{m}^\dagger/\sqrt{2}${, where $L_{m,\pm}$ are  functions of $c_m$ and $c_m^\dagger$ only.}
{In case when different $c_m$'s correspond to separate sites $m$ then these Lindblad operators become fully local.}

If $t-t_0$ is large compared to the characteristic timescale
\begin{equation} \label{tau}
     \tau_\text{c} = \min \left( \frac{1}{\kB \T}, \frac{1}{|\D E - \mu|} \right)
\end{equation}
given by the smaller of the inverse thermal energy and the inverse energy distance to the chemical potential, the integral \eqref{ReF} converges and the coefficients 
\begin{equation} \label{ReF-latetime}
  F^\pm_{\D E, \T}(t\gg \tau_\text{c}) \approx f_\pm(\D E, \T)
\end{equation}
approach the Fermi function for the bath temperature $\T$. 

Our main finding, 
from the technical point of view, is that the integral in \eqref{ReF} can be uniformly in $t, \D E$ and $\T$ approximated by the function
\begin{align} \label{int}
    I &= \kB \T \int_{0}^{t-t_0} \frac{\sin[(\D E - \mu)\tau]}{ \sinh(\pi \kB \T\tau)} d\tau \\ &\approx \frac{1}{\pi} \Si\left[\frac{\pi (\D E - \mu)}{4 \kB \T} \tanh\left( \frac{4 \kB \T (t-t_0)}{\pi} \right) \right], \label{I_SI_tanh}
\end{align}
with $\Si$ being the sine integral, $\Si(x) = \int_0^x \sin(u)/u\, du$. 
It relies on the astonishing similarity \eqref{magic} discussed in App.~\ref{app:Similarity} (cf. Fig.~\ref{fig:Similarity})
and offers a good approximation when $\T (t-t_0) \lesssim \pi/4$.
%
It still shows the unwanted excess oscillations as in original $F^\pm_{\D E, \T}(t)$ (cf. Fig.~\ref{fig:F(DE)}). 
{These can be most clearly observed in the limit $\T\ra 0$ when \eqref{I_SI_tanh} becomes exact
\begin{equation} \label{int-approx_T=0}
    I = \int_{0}^{t-t_0}\text{d}\tau \frac{\sin[(\D E - \mu)\tau]}{ \pi \tau} 
    = \frac{\mbox{Si}[(\D E - \mu)(t-t_0)]}{\pi}
\end{equation}
and can assume values out of the range $[-\frac{1}{2}, \frac{1}{2}]$ which may lead to negative values of $F^\pm_{\D E, \T}(t)$ in \eqref{ReF}.}
Therefore, in the last step, we replace the sine integral function by $\tanh$ which has a similar form but stays bounded in the proper region without any oscillations (cf. Fig. \ref{fig:SinInt-Tanh}), by which we arrive at
\begin{align} \label{int-approx}
    I &\approx 
    \frac{1}{2} \tanh\left[ \frac{\D E - \mu}{2\, \kB \T} \tanh\left( \frac{4\, \kB \T (t-t_0)}{\pi} \right) \right] 
\end{align}
(cf. App.~\ref{app:TeffDeriv2} for the full derivation).

\begin{figure}[t]
\includegraphics[width=\linewidth]{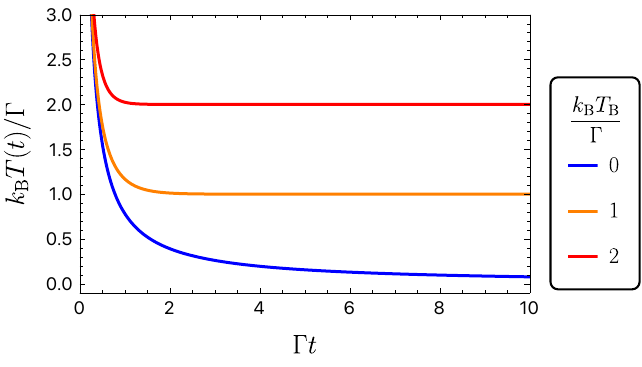}
  \caption{Evolution of the time--dependent effective temperature $T(t)$ for various temperatures of the bath $\T$.}
  \label{fig:T(t)}
\end{figure}

In the limit $t\ra\infty$, the approximation \eqref{int-approx} becomes an equality (with the inner $\tanh(\infty) = 1$) and inserted into \eqref{ReF} delivers%
\footnote{Due to the identity $\frac{1}{2} \mp \frac{1}{2} \tanh[(\D E - \mu)/(2\kB \T)] = f_\pm(E,\T)$.}
the static Fermi function \eqref{ReF-latetime}.
By inserting the full approximation obtained in \eqref{int-approx} into the formula \eqref{ReF} 
and denoting the approximated coefficients as $\F^\pm_{\D E, \T}(t)$
we find that the result can be recast as a Fermi function, too, (cf. Fig. \ref{fig:F(DE)}, top)
\begin{equation} \label{ReF_eff_T}
    F^\pm_{\D E, \T}(t) \approx \F^\pm_{\D E, \T}(t) \equiv F^\pm_{\D E, T(t)}(\infty) = f_\pm(\D E, T(t)),
\end{equation}
now with a modified, time--dependent temperature 
\begin{equation} \label{Teff}
    T(t) = \T \Bigl/\tanh\left(\frac{4\kB \T}{\pi}(t-t_0)\right),
\end{equation}
(cf. Fig.~\ref{fig:T(t)}).
The effective temperature $T(t)$ is universal, i.e. independent of $\D E$. 
It diverges at short times,
\begin{equation} \label{T_eff_0}
  T(t) \approx  \frac{\pi}{4\kB(t-t_0)} \equiv T_0(t) 
  \quad \text{for} \quad t - t_0 \ll \frac{1}{\kB \T}
\end{equation}
and converges to the true bath temperature $T(t) \ra \T$ for late times $t \ra \infty$ (cf. Fig.~\ref{fig:T(t)}).
We will refer to it as the time--dependent effective bath temperature, for it is described by the effective bath temperature from the perspective of the system.
The origin of its time--dependence is in the contact of the system with the environment at $t=t_0$ and the build--up of correlations (which were assumed to be absent for $t<t_0$). 
It originates in short time off--resonant (virtual) transition processes between the bath and the system involving an energy range close to the Fermi--level, proportional to the inverse of the characteristic time of the associated oscillations, which scales as $(t-t_0)^{-1}$, staying independent of other parameters, such as $\Ga$ or $\T$.
Also for later times, $T(t)$ cannot depend on the coupling strength $\Ga$ because we keep only the lowest order terms in \eqref{vonNeumann} and hence $\Ga$ automatically factorizes in \eqref{F(t)} and \eqref{ReA} and plays only the role of a scaling factor.

The approximate coefficients directly
satisfy three important limiting cases, for $t\ra \infty, t \ra 0$ and $\T\ra \infty$, which can be directly obtained from the integral \eqref{ReF}
and the fourth limiting case, for $\T\ra 0$, which resolves the problem of excess oscillations observed in \eqref{int-approx_T=0} and replaces the result with a non--oscillatory function,
\begin{align}
  \F^\pm_{\D E, \T}(t\ra \infty) &= \frac{1}{2} \mp \frac{1}{2} \tanh \frac{\D E - \mu}{2 \kB \T} = f_\pm(\D E, \T), \nonumber \\
  \F^\pm_{\D E, \T}(t \ra 0) &= \frac{1}{2} = f_\pm(\D E, \infty), \nonumber \\
  \F^\pm_{\D E, \T \ra \infty}(t) &= \frac{1}{2} = f_\pm(\D E, \infty), \nonumber \\
  \F^\pm_{\D E, \T \ra 0}(t) &= \frac{1}{2} \mp \frac{1}{2} \tanh\!\frac{2 (\D E - \mu) (t-t_0)}{\pi } \nonumber \\
  &= f_\pm(\D E, T_0(t)),
\end{align}
where $T_0(t)$ is given in \eqref{T_eff_0}. 
In all cases $\F^\pm_{\D E, \T}(t)$ stays in the required range $[0,1]$ (cf. dashed lines in Fig.~\ref{fig:F(DE)}, top). 

\begin{table*}
  \setlength{\tabcolsep}{8pt}
  \renewcommand{\arraystretch}{1.7}
  \begin{tabular}{|c|c|c|c|} \hline
     Approach & Coefficients &  Value & for $\T=0$ \\ \hline
     Redfield--I & $F^\pm_{\D E, \T}(t)$ & & $\frac{1}{2} \mp \mbox{Si}(\D E t)/\pi$ \\ \hline
     modified Redfield--I & $\F^\pm_{\D E, \T}(t) = f_\pm(\D E, T(t))$ & $\frac{1}{2} \mp \frac{1}{2} \tanh\left[ \frac{\D E}{2\,\kB \T} \tanh\left( \frac{4\,\kB \T t}{\pi} \right) \right]$ & $\frac{1}{2} \mp \frac{1}{2}\tanh\left(\frac{2 \D E t}{\pi}\right)$ \\ \hline
     static Redfield--II & $F^\pm_{\D E, \T}(\infty) = f_\pm(\D E, \T)$ & $\frac{1}{2} \mp \frac{1}{2} \tanh\left( \frac{\D E}{2\,\kB \T} \right)$ & $\Theta(-\D E)$\\ \hline
  \end{tabular}
  \caption{Overview of the discussed approximation schemes. The constants $\mu$ and $t_0$ were set to zero for brevity.}
  \label{tab:approaches}
\end{table*}

\subsection{Imaginary part of $A^\pm_m(t,\D E)$}

After we have considered the real part of $A^\pm_m(t,\D E)$ we next focus on its imaginary part. In the wideband limit, $\text{Im}\,A^\pm_m(t, \D E)$ diverges. However, \eqref{H-ren} and \eqref{M=F+F} contain only differences, $\text{Im}\,A^\pm_m(t, \D E)-\text{Im}\,A^\pm_m(t, \D E')$, which are finite. Therefore we define
\begin{equation} \label{ImA}
    \Ga_m G_{\D E, \T}(t)= \Im\,A^\pm_m(t,\D E)-\Im \,A^\pm_m(t,\mu)
\end{equation}
{(which is independent of $\pm$)}
by choosing a universal value, $\D E' = \mu$, for the counter--term and obtain
\begin{multline} \label{ImF}
    G_{\D E, \T}(t) \equiv 
   \kB \T\int_{0}^{t-t_0} \frac{1-\cos[(\D E - \mu)\tau]}{ \sinh(\pi \kB \T\tau)} d\tau.
\end{multline}
Applying similar approximations as for the real part of $A^\pm_m(t,\D E)$ 
we find
\begin{equation} \label{G-approx}
    G_{\Delta E,\T}(t)\approx \frac{1}{\pi}\text{Re}\Biggl[\psi\left(\frac{1}{2}+i \frac{\D E-\mu}{2 \pi \kB T(t)} \right)-\psi\left(\frac{1}{2}\right)\Biggr]
\end{equation}
with the same effective temperature $T(t)$ as in \eqref{Teff} (cf. Fig.~\ref{fig:F(DE)} and App.~\ref{app:TeffDeriv3}) and $\psi(x)$ the digamma function \cite[Sec. 6.3]{abramowitz1968handbook}.  
Denoting the right--hand side by $\G_{\Delta E, \T}(t)$, 
we arrive, in full analogy to \eqref{ReF_eff_T}, at
\begin{equation} \label{ImF_eff_T}
    G_{\D E, \T}(t)\approx \G_{\Delta E, \T}(t) \equiv G_{\D E, T(t)}(\infty).
\end{equation}
This confirms that the same effective temperature $T(t)$ can be used in both, the real and the imaginary part of $A^\pm_m(t,\D E)$.

{In the next two sections, \ref{sec:singleDot} and \ref{sec:benchmark}, we will consider a simple system consisting of a single quantum dot in order to compare the different  approaches analytically and numerically: 
\begin{itemize}[leftmargin=16pt, topsep=0pt, itemsep=1pt, partopsep=0pt, parsep=0pt, after=\vspace{0pt}]  
  \item exact solutions (for the non--interacting system),
  \item Redfield--I equation with time--dependent coefficients $F^\pm_{\D E, \T}(t)$ leading to positivity problem \textbf{1} due to overshooting the range $[0,1]$,
  \item modified Redfield--I equation with approximate coefficients $\F^\pm_{\D E, \T}(t) \in [0, 1]$ based on time--dependent temperature $T(t)$,
  \item Redfield--II equation with static coefficients $F^\pm_{\D E, \T}(\infty) \in [0, 1],$
\end{itemize}
summarized in Tab.~\ref{tab:approaches}.
{We will concentrate only on the real part $F^\pm_{\D E,\T}(t)$ since the imaginary part $G_{\D E, \T}(t)$ will not be significant for that system.}
For larger systems than one quantum dot, we would also deal with the problem \textbf{2} of non--positivity of some $2 \times 2$ matrix valued coefficients, $\M \ngeq 0$, in which case we might want to further approximate the Redfield--I and II equations to the Lindblad form with modified $\t\M \geq 0$ (cf. App. \ref{app:RedToLind}).
}

\section{Example: single Quantum dot} \label{sec:singleDot}

Here, we consider a simple system to demonstrate the application of the above proposed approximation method based on the time--dependent temperature. 
We choose a single quantum dot described by 
the Anderson impurity model \cite{Andersonmodel}
\begin{equation} \label{Ham-1dot}
    H_\text{S}=U n_\uparrow n_\downarrow + \eps (n_\uparrow+n_\downarrow)
\end{equation}
with $n_s=c^\dagger_s c_s$, spin $s=\uparrow,\, \downarrow$, Coulomb interaction $U>0$ and onsite energy $\eps$, satisfying  
$-U < \eps - \mu < 0$. 
{The system has $\dim(S)=4$ eigenstates, cf. \eqref{HSeigen}, and is connected to $M=2$ baths with different spin polarizations, cf. \eqref{HBintro} and \eqref{HCintro}.}
Both, in the static (second) Markov approximation, for $t_0\rightarrow -\infty$, as well as in the 
time--dependent effective temperature approximation, with $t_0 = 0$, \eqref{Redfield-mod} reduces to a {GKSL} equation (cf. App. \ref{app:RedToLind}) with the Lindblad dissipators \eqref{Lind-glob-sec} 
or \eqref{Lind-glob-coh} 
{(which differ only by non--physical coherences between states with different occupation numbers)}
\begin{align} \label{L-loc1}
     L^\pm_{s,1} &= \sqrt{f_\pm(\eps, \t T(t))}\, c^\pm_s (1-n_{\bar s}), \\
     L^\pm_{s,2} &= \sqrt{f_\pm(\eps+U, \t T(t))}\, c^\pm_s n_{\bar s}, \label{L-loc2}
\end{align}
for tunneling in/out ($\pm$) of the first \eqref{L-loc1} and second electron \eqref{L-loc2} with spin $s$.
The temperature $\t T(t)$ in the Fermi functions $f_\pm$ is either constant and equal $\T$ or time--dependent as given by \eqref{Teff}. 
In the static case, for $\T=0$, the states $\ketbra{\uparrow}{\uparrow}$ and $\ketbra{\downarrow}{\downarrow}$ are ``frozen''. 
However, the time--dependent effective temperature $T(t)$, even for $\T=0$, becomes initially large, $T(t\approx 0) \ra \infty$, cf. \eqref{T_eff_0}, and the system is temporarily driven towards the fully mixed ``hot'' state
\begin{equation}
    \rho_\infty = \frac{1}{4}\Bigl[\ketbra{0}{0} + \ketbra{\su}{\su} +\ketbra{\sd}{\sd} + \ketbra{\su\sd}{\su\sd}\Bigr].
\end{equation}
Eventually, as $T(t) \ra \T=0$, it relaxes to some mixture $a \ketbra{\su}{\su} + b \ketbra{\sd}{\sd}$ with $a+b=1$.
Starting with the pure state $\ketbra{\su}{\su}$, 
the final spin--$z$ $\ev{S_z} = (a - b)/2$
will measure how strong the influence of the ``hot'' period on the effective dynamics was.
This observable satisfies an autonomous differential equation%
\footnote{If $\t \L_t^\dagger$ is time--dependent, $\dot{A}_\text{H}(t)=\t \L_t^\dagger A_\text{H}(t)$ is generally not correct for any operator $A_\text{H}$ in the Heisenberg picture. However, if the equation is autonomous then it holds true.}
(valid for any $\T$)
\begin{equation}
    \dot{S}^\text{eff}_z(t) = - \Ga \Bigl[1-f_+(\eps,T(t))+f_+(\eps+U,T(t))\Bigr] S^\text{eff}_z(t)
\end{equation}
which can be integrated to
\begin{equation}
    S^\text{eff}_z(t) = S_z(0) \exp\left[-2 \Ga \int_0^t \text{d}t' f_+(\eps+U,T(t')) \right]
    \label{Szteff}
\end{equation}
where we have chosen the special value $\eps = \mu-U/2$ for convenience. For $\T>0$ this integral is difficult to calculate but it diverges for $t\ra\infty$ and thus leads to $S_z(\infty) = 0$. 
For $\T=0$, we have $\kB T(t) = \pi/(4 t)$ (cf. \eqref{Teff}) and this integral can be calculated exactly to give
\begin{equation}
     S^\text{eff}_z(t) = S_z(0)\, e^{-2\Ga t}\left(1+e^{\frac{2U}{\pi}t}\right)^{\frac{\Gamma \pi}{U}} 2^{-\pi \frac{\Gamma}{U}}
\end{equation}
which for $t\gg 1/U$ has the limit
\begin{equation} \label{Sz(inf)-Teff}
      S^\text{eff}_z(t) \cong  2^{-\pi \Ga/U} S_z(0).
\end{equation}
It means that the initial spin--z, $S_z(0)$, decays in a non--perturbative way, which is enhanced by the system--bath coupling $\Ga$ and suppressed by the Coulomb repulsion $U$. 
This is in contradiction to the observation that both pure states are ``frozen'' in the static (second) Markov case. 

\begin{figure}[ht!]
\includegraphics[width=\linewidth]{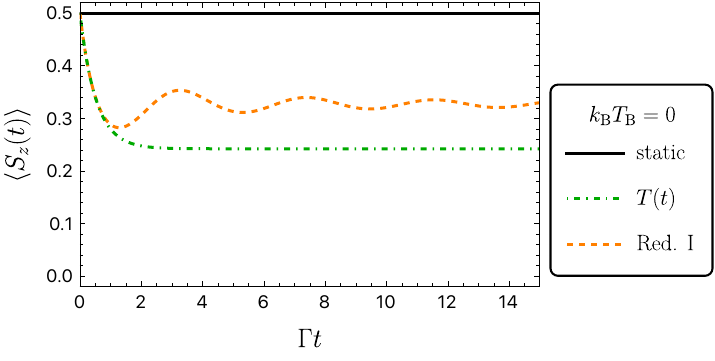}
\includegraphics[width=\linewidth]{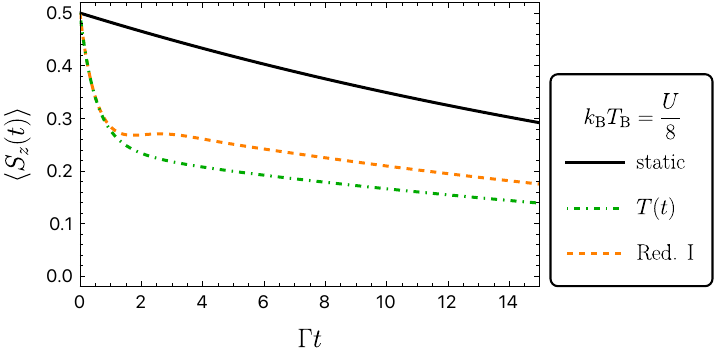}
\includegraphics[width=\linewidth]{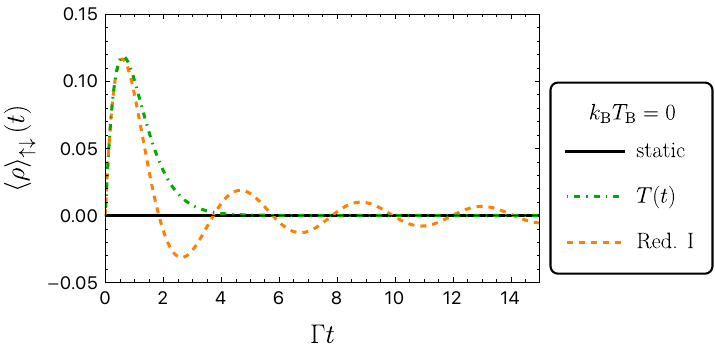}
\includegraphics[width=\linewidth]{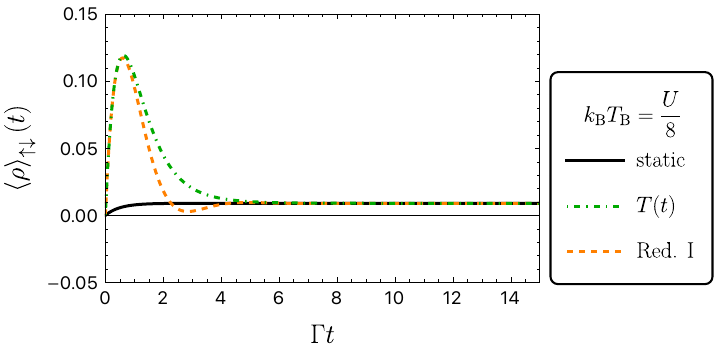}
  \caption{One quantum dot. Top pair: spin $\ev{S_z(t)}$ as a function of time for the static Markov and the effective temperature approximations as well as for the Redfield--I master equation \eqref{Redfield-mod}. 
  The Coulomb interaction is $U=3\,\Gamma$, 
  the onsite potential is $\eps=\mu-\frac{U}{2}$,
  and the bath temperature is $\T=0$ or $\kB \T=U/8$.
  The initial state is $\rho_S(0) = \ketbra{\su}{\su}$. 
  Bottom pair: Probability of the double occupation $\ev{\rho}_{\su\sd}(t) \equiv \braket{\su \sd|\rho(t)|\su \sd}$ as a function of time, here identical with the zero occupation $\ev{\rho}_0(t) \equiv \braket{0|\rho(t)|0}$.
  Both can get negative for the Redfield equation.
  The energy, given by $\ev{H_\text{S}-\mu N}=U\left[\ev{\rho}_{\su\sd}+\ev{\rho}_{0}-1\right]/2$ with $N=n_\su+n_\sd$, also swings below its theoretical lower limit $-U/2$ for $\T=0$.  
  The parameters are the same as for $\ev{S_z(t)}$, respectively.}
  \label{fig:1DotSz}
\end{figure}

Also for the not approximated Redfield--I master equation \eqref{Redfield-mod} the spin--$z$ can be calculated analogously to \eqref{Szteff}
\begin{equation}
    S_z(t) = S_z(0) \exp\left[-2\, \Ga \int_0^t F^+_{\D E = \eps+U, \T}(t')\, dt' \right].
    \label{SzRed}
\end{equation}
In this particular system and for $\T=0$ 
the integral can be evaluated exactly and gives
\begin{multline}
    S_z(t) = S_z(0)\, e^{-\Gamma t } \times \\ 
    \exp \Biggl[ \frac{4\, \Gamma}{\pi U}\left(\cos\left(\frac{Ut}{2}\right)-1\right)
    + \frac{2\,\Gamma\, t}{\pi} \Si\left(\frac{Ut}{2}\right)\Biggr]
\end{multline}
which for $t\gg 1/U$ gives 
\begin{equation} \label{Sz(inf)-Red}
     S_z(t) \cong  e^{-4\,\Gamma/(\pi U)} S_z(0).
\end{equation}
Due to $2^{-\pi \Ga/U}<e^{-4\,\Gamma/(\pi U)}$, the approximation with the effective temperature \eqref{Sz(inf)-Teff} slightly overestimates the effect compared to the prediction of the Redfield--I master equation \eqref{Sz(inf)-Red}, however, both stay within the same order of magnitude which justifies the effective temperature approximation (cf. Fig.~\ref{fig:1DotSz}). 
On the other hand, the Redfield--I master equation \eqref{Redfield-mod} does not preserve the positivity of the density matrix (cf. Sec.~\ref{sec:TimeDepCoef}) and leads to negative probability, as shown in Fig.~\ref{fig:1DotSz} (second from bottom) 
whereas the effective temperature approximation is free of this problem. 
For bath temperatures $\T>0$, both results decay to $S_z(\infty)=0$ at late times. 
However, at short times, $t\ll \tau_\text{c}$, and small bath temperature $\T$, the Redfield--I master equation and the effective temperature approximation give $S_z(t) \sim \exp\left(-\Gamma t\right)$ while in the static approximation the decay is exponentially suppressed, $S_z(t) \sim  \exp\left(-2\,\Gamma e^{-U/(2 \kB \T)}t\right)$.

\begin{figure}[t]
\includegraphics[width=\linewidth]{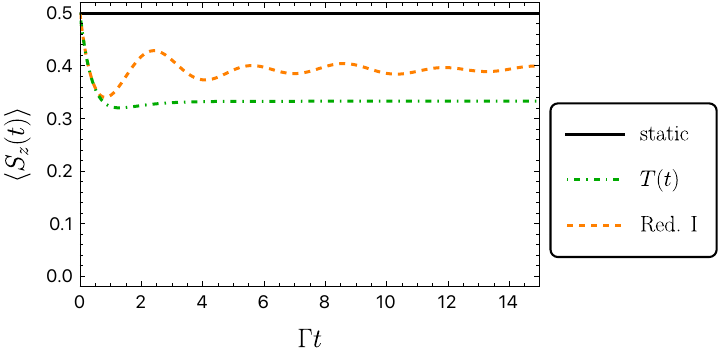}
\includegraphics[width=\linewidth]{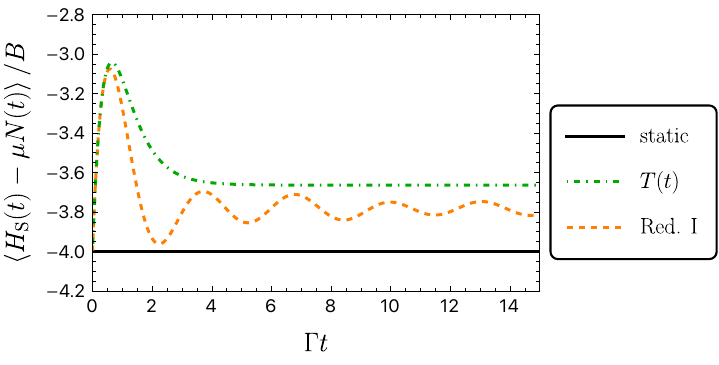}
  \caption{One quantum dot. Spin $\ev{S_z(t)}$ (top) and energy $\ev{H_\text{S}(t)-\mu N(t)}$ with $N=n_\su+n_\sd$ (bottom) as  functions of time for the static Markov and the effective temperature approximations as well as for the Redfield--I master equation (\ref{Redfield-mod}). The Coulomb interaction is $U=3\,\Gamma$, the onsite potential is $\eps=\mu-U/2$, the bath temperature is $\T=0$, the magnetic field is $B=-\Gamma/2$, 
  and the initial state is $\rho_S(0) = \ketbra{\su}{\su}$ (ground state).
  }
  \label{fig:1DotSzmitB}
\end{figure}

{By a similar mechanism, the effective temperature can even lead to an excitation of the ground state and deposit energy into the system.}
By adding a magnetic field in the $z$--direction with the Hamiltonian 
$ H_z =  2 B S_z $
we obtain together with \eqref{Ham-1dot} 
\begin{equation}
  H_S = U n_\su n_\sd + (\eps + B) n_\su + (\eps - B) n_\sd
\end{equation}
and lift the degeneracy between the $\ket{\uparrow}$ and $\ket{\downarrow}$ states.
We choose $\T=0$ and start in the ground state $\rho_S(0) = \ketbra{\su}{\su}$ (for $B<0$).
By connecting the system with the bath at $t = t_0 = 0$, due to the high effective temperature $T(t)$ at short times, there will be an increase of energy in the system 
\begin{equation}
    \D H_\text{S}=H_\text{S}(\infty)-H_\text{S}(0)=2B(S_z(\infty)-S_z(0)),
\end{equation}
cf. Fig.~\ref{fig:1DotSzmitB}. 
Although the bath and the system were initially in their respective ground states, the increase of energy, $\Delta H_S$, results from the coupling Hamiltonian, $H_\text{C}$. In the limit of weak $B$, it can be evaluated to
\begin{equation}
    \D H_\text{S}=|B|\left(1- e^{-4\,\Gamma/(\pi U)}\right) + \mathcal{O}\left(\frac{B^2}{U}\right)
\end{equation}
for the Redfield--I master equation (\ref{Redfield-mod}) and to
\begin{equation}
    \D H_\text{S}=|B|\left(1- 2^{-\pi \Gamma/U}\right) + \mathcal{O}\left(\frac{B^2}{U}\right)
\end{equation}
for the effective temperature approximation,
where again the latter method slightly overestimates the result. 

\section{Non--interacting quantum dot as benchmark} \label{sec:benchmark}

Since the proposed effective temperature method is an approximation to the Redfield equation which in turn is also an approximation itself, it does not provide a proper benchmark for testing the accuracy. Especially in the situations when the Redfield equation leads to mathematical problems the comparison is unclear. 
Therefore, we consider here the non--interacting case with $U=0$ which is exactly solvable 
and compare the different master equation approaches with it.

Without the Coulomb interaction, the system splits into two identical copies of a spinless system 
\begin{equation}
    H = \eps\, n + \sum_{k}\left(\gamma_{k}c^\dagger\,b_{k} + \text{h.c.}\right) + \sum_{k}\eps_{k}b_{k}^\dagger b_{k}
    \label{HU=0}
\end{equation}
{with $n=c^\dagger c$.}
Starting from the Heisenberg equation of motion and  Laplace transformation technique \cite{Topp_2015,JussiauPhysRevB.100.115411}, it is possible to {express the annihilation operator of the dot in the Heisenberg picture}
\begin{equation}
    c_\text{H}(t)=e^{-i\eps t -\frac{\Gamma}{2}t} c_\text{H}(0) +\sum_k \gamma_k b_k \left(\frac{e^{-i\eps t -\frac{\Gamma}{2}t}-e^{-i\eps_k t}}{(\eps-\eps_k)-i\frac{\Gamma}{2}}\right)
    \label{cexact}
\end{equation}
in terms of the operators $c = c_\text{H}(0)$ and $b_k$ in the Schrödinger picture. 
Assuming an initial product state between the dot and a Fermi distributed bath leads to 
\begin{multline}
    \left<n(t)\right>=e^{-\Gamma t}\left<n(0)\right>\\+\Gamma \int_{-\infty}^\infty \frac{\text{d}\omega}{2 \pi}f_+(\omega+\eps,\T) \frac{e^{-\Gamma t}+1-2 \cos(\omega t)e^{-\frac{\Gamma}{2}t}}{\omega^2+\frac{\Gamma^2}{4}}.
    \label{exactn(t)}
\end{multline}

\begin{figure}[t]
\includegraphics[width=\linewidth]{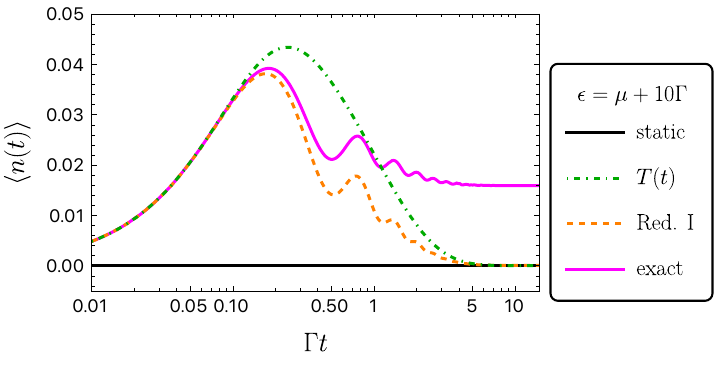}
\includegraphics[width=\linewidth]{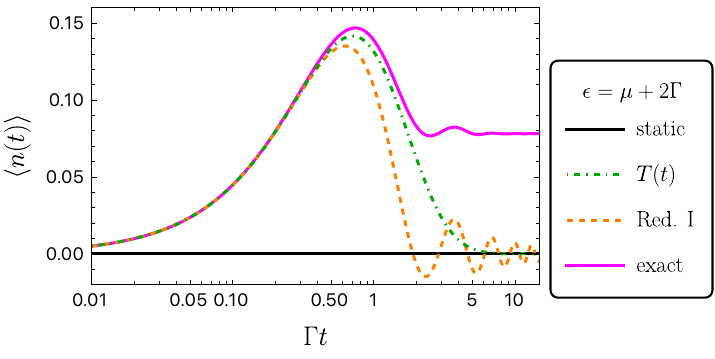}
	\caption{Particle number $\left<n(t)\right>$  as a function of time (logarithmic scale). The lines show the various master equations and exact solution (\ref{exactn(t)}) with $\T=0$, $U=0$, $\ev{n(0)}=0$ and $\eps=\mu+10\,\Gamma$ (top) or $\eps=\mu+2\,\Gamma$ (bottom).}
 \label{fig: exactvsRedfield}
\end{figure} 

\begin{figure*}[t]
\includegraphics[width=\linewidth]{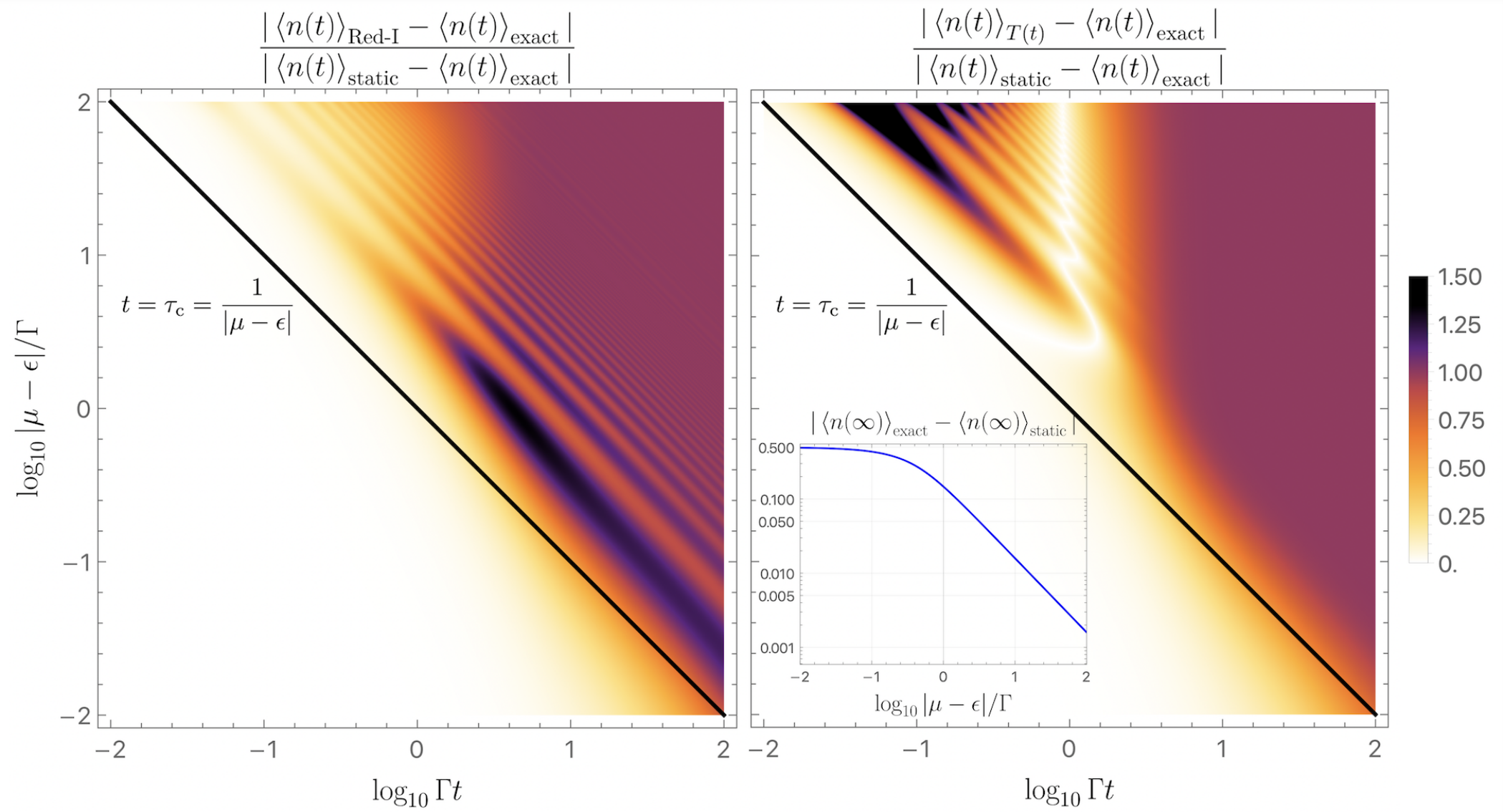}
  \caption{The normalized difference of the particle number between the exact solution (\ref{exactn(t)}) and the Redfield--I equation (left panel) or the Redfield--I equation modified with $T(t)$ (right panel) as a function of time and of the difference between the chemical potential $\mu$ and the onsite potential $\eps$ (both in logarithmic scales). The normalization is given by the difference  between the exact and the static result. For both situations (left and right), the normalized and non--normalized deviations are small if the time is smaller than the characteristic time scale $\tau_\text{c}$. The black line shows the correspondence between $\tau_\text{c}$ and $|\mu-\eps|$. The inset shows the non--normalized deviation between the particle number of the steady state of the static Redfield master equation and the exact solution as a function of time. This value decays as $\left|\left<n(\infty)\right>_\text{exact}-\left<n(\infty)\right>_\text{static}\right| \approx \Gamma/(2 \pi |\mu-\eps|)$ and vanishes in the weak coupling limit. The system is described by \eqref{HU=0} with bath temperature $\T=0$.}
  \label{fig: n_exactvsRedfield_2D}
\end{figure*}

Fig.~\ref{fig: exactvsRedfield} compares the exact solution \eqref{exactn(t)} and the different master equations as functions of time. Due to forbidden energy transitions for $\eps>\mu$ and $\T=0$, the static approximation leads to no dynamics when starting with an initially empty dot
while in all other approaches we observe relaxation at short times, $t\lesssim \tau_\text{c}$.
Similarly to Fig.~\ref{fig:1DotSzmitB}, in Fig.~\ref{fig: exactvsRedfield} we observe an energy gain $(\eps -\mu)(\left<n(t)\right>-\left<n(0)\right>)$ in the system which originates from the coupling term in the Hamiltonian, $H_\text{C}$ (in App.~\ref{app:exactCorrelationswithbath}, we discuss this in more detail).
By comparing the exact solution with the results of the different master equations, we find that both time--dependent Redfield--I master equations, {original and modified with the effective temperature,} are accurate for times shorter than the characteristic timescale $\tau_\text{c}$ \eqref{tau} (cf. Fig.~\ref{fig: n_exactvsRedfield_2D}).
The accuracy for short times holds still true even in the case of strong coupling, $\Gamma \gtrsim |\mu-\eps|$, where the exact solutions show significant deviations from all Redfield master equations at late times, $t\gg \tau_\text{c}$
(cf. Fig.~\ref{fig: n_exactvsRedfield_2D}).

In the regime of intermediate times and strong coupling $\Gamma$, the original Redfield--I master equation may lead to non--physical values of the particle number below zero or above one (cf. Fig.~\ref{fig: exactvsRedfield} (bottom))
where the Redfield--I master equation modified with $T(t)$ is superior (cf. Fig.~\ref{fig: n_exactvsRedfield_2D}).
The opposite happens for
weak coupling, $\Gamma \ll |\mu-\eps|$ (cf. Fig.~\ref{fig: exactvsRedfield} (top) and Fig.~\ref{fig: n_exactvsRedfield_2D})
where the effective temperature method overshoots, as discussed in Sec.~\ref{sec:singleDot}.
However, the differences between all master equations and the exact solution vanish in that regime as $\Gamma \tau_\text{c} = \frac{\Gamma}{|\mu-\eps|}\rightarrow 0$ (cf. Fig.~\ref{fig: n_exactvsRedfield_2D} (inset)).

The above considerations extend qualitatively to the regime $k_\text{B}\T \ll \frac{|\mu-\eps|}{4}$
where the largest deviations between different approaches are expected.
In contrary, for temperatures $k_\text{B}\T\gtrsim \frac{|\mu-\eps|}{4}$, the original Redfield--I master equation no longer reaches non--physical values of the particle number and becomes similar to the modified Redfield--I master equation. Analogously to $\T=0$, the differences between the master equations and the exact solution vanish in the regime of high $\T$ as $\Gamma \tau_\text{c} = \frac{\Gamma}{k_\text{B}\T} \rightarrow 0$.


\section{Conclusions}

We considered a fermionic quantum system exchanging particles with a thermal bath at fixed temperature $\T$. 
By eliminating the bath from the description we derived an approximation of the first Redfield equation with time--dependent coefficients which offers the interpretation of a time--dependent effective temperature $T(t)$ \eqref{Teff}. 
For times smaller than the characteristic time $\tau_\text{c}$ \eqref{tau}, 
the effective temperature is very large,  $T(t) \ra \infty$, 
which can be explained by the fact that at short timescales all energetically forbidden transitions are allowed%
\footnote{{An entirely different approach \cite{Wegewijs-Renormalization}, based on the renormalization flow in the environment's temperature, demonstrated that the short--time dynamics of observables shows a universal temperature--independent behavior when the metallic reservoirs have a flat wide band. This is in a perfect agreement with our observation of infinite effective temperature which makes the renormalization flow trivial.}}.
At timescales much larger than $\tau_\text{c}$, the effective temperature $T(t)$ converges against the true environment temperature $\T$. The use of the effective temperature $T(t)$ fixes also the problems of non--positivity in the development of the density matrix by the Redfield equation. 

We demonstrated these effects on the example of a single quantum dot with Coulomb interaction (Sec.~\ref{sec:singleDot}) which we also compared with exact solutions for a system without the Coulomb interaction, using it as a benchmark (Sec.~\ref{sec:benchmark}).
In both cases we have shown a qualitative and quantitative agreement between the solutions of the original time--dependent Redfield master equation and its approximation based on the time--dependent temperature. 
In the case without the Coulomb interaction, we have shown a good agreement between the solutions of the Redfield equations and the exact solution. 
For short times, we see a perfect agreement even for parameters for which the Redfield approximation generally does not hold at later times.
This confirms that the effective temperature scheme offers an approximation which is as close to the exact solution as the original Redfield equation and combines a perfect match at short times with being free of any mathematical flaws at later times.

Previous versions of time-dependent effective temperatures discussed in the literature \cite{hebel1959nuclear,Primakoff_PhysRev.130.1267,andersen1964exact, T(t)systemPhysRevA.64.052110(2001), Schaller_2014,ME_mesoReserviorPhysRevE.105.054119(2022),mesoReserviorPhysRevLett.131.220405(2023)}
differ substantially from the approach presented here. 
Most of them are related to time-dependent quasi--equilibrium temperatures of the system or of the bath, based on their internal dynamics. 
In contrast, in our approach the bath stays in thermal equilibrium at $\T$ and the \textit{effective bath temperature} $T(t)$ refers to its observed value from the perspective of the system, emerging as a consequence of switching on the interaction and buildup of correlations between the system and the bath.
Our approach is based on a sophisticated analytic approximation holding for a wide range of temperatures $\T$, energies $\D E$ and time--scales $t$, \eqref{int-approx}--\eqref{Teff},
and agrees with the Redfield--I equation when $t-t_0 \lesssim \tau_\text{c}$ \eqref{tau}.
Despite this, it preserves the original structure of the standard static results \eqref{ReF_eff_T}, replacing only $\T$ with $T(t)$ which, among others, can be practical in numerical simulations. 


The method of time--dependent effective bath temperature
lefts open space for further approximations and, therefore, can be applied to various types of originally time--independent master equations, beyond the secular approximation or time--averaging methods.
Its potential application to bosonic environments remains open due to the qualitative differences between the Bose and the Fermi statistics.

\section{Acknowledgments}

The authors thank Gernot Schaller for fruitful discussions and valuable feedback on the manuscript.
We gratefully acknowledge funding by the Deutsche Forschungs\-gemeinschaft (DFG, German Research Foundation) --- Project 278162697 --- SFB 1242.

\appendix
\section{From Redfield to Lindblad}\label{app:RedToLind}

In the derivation of the Redfield--I equation \eqref{Redfied-I} (cf. Sec.~\ref{sec:TimeDepRed}), at the lowest order in the system--environment coupling strength, 
the preservation of positivity of the density matrix in the evolution gets lost. 
Here, we discuss its corrections leading towards GKSL master equations 
\cite{Kirsanskas+Wacker-CohLindPhenom, Davidovic2020, Nathan-CohLindDerivation, B07-Dimer, B07-Trimer,Breuer, Davies1}.

For transition processes in the system involving the energy differences, $\D E$ and $\D E'$, we consider the blocks
\begin{widetext}
\begin{equation} \label{Redfield-matrix}
\begin{split}
  &\M^\pm_{\D E, \D E'}(t)  = 
  \begin{pmatrix}
    M_\pm(t,\D E, \D E), & M_\pm(t,\D E, \D E') \\
    M_\pm(t,\D E', \D E), & M_\pm(t,\D E', \D E')
  \end{pmatrix} \\
  &=
  \begin{pmatrix}
    F^\pm_{\D E, \T}(t), & 
   \frac{1}{2} [F^\pm_{\D E, \T}(t) + F^\pm_{\D E', \T}(t)] + \frac{i}{2} [G_{\D E, \T}(t) - G_{\D E', \T}(t)] \\
    \frac{1}{2} [F^\pm_{\D E, \T}(t) + F^\pm_{\D E', \T}(t)] - \frac{i}{2} [G_{\D E, \T}(t) - G_{\D E', \T}(t)], & 
    F^\pm_{\D E', \T}(t)
  \end{pmatrix}
\end{split}
\end{equation}
\end{widetext}
built with $M_\pm(t,\D E, \D E') \equiv M_m^\pm (t,\D E, \D E')/\Gamma_m$ independent of $m$
(the coefficients $F$ and $G$ are defined in \eqref{ReF} and \eqref{ImF}, respectively). 
If the coefficients $F^\pm_{\D E, \T}(t)$ become negative (cf. Fig.~\ref{fig:F(DE)}) the blocks $\M$ cease to be positive definite that translates also into the total Liouville operator \eqref{LL}. 
Therefore, we propose in this work the approximation \eqref{ReF_eff_T} which replaces them with new coefficients $\F^\pm_{\D E, \T}(t) \in [0,1]$.
For consistency, we also replace the $G^\pm_{\D E, \T}(t)$ coefficients with $\G^\pm_{\D E, \T}(t)$ according to \eqref{ImF_eff_T} (cf. Sec.~\ref{sec:TimeDepCoef}).

However, in general, also the approximated matrices $\mathcal{M}^\pm_{\D E, \D E'}(t)$ are not positive definite for $\D E \neq \D E'$, having one positive and one negative eigenvalue,
which presents another possible reason for the the non-positivity of the total Liouville operator \eqref{LL}.

\subsection{Secular approximation}

The off--diagonal elements of $\M^\pm_{\D E, \D E'}(t)$ correspond to coherences between states with different energies (in the Liouville space), which oscillate in time. 
The secular approximation \cite{Breuer, Davies1} averages out the oscillations and effectively removes the off--diagonal terms by which 
\begin{align}
  \mathcal{M}^{\pm,\text{sec}}_{\D E, \D E'}(t) = 
  \begin{pmatrix}
    \F^\pm_{\D E, \T}(t) & 0 \\
    0 & \F^\pm_{\D E', \T}(t)
  \end{pmatrix}
\end{align}
are obviously positive definite. 
It is equivalent to the replacement of the coefficients $M^\pm_{m,s}$ in \eqref{M=F+F} with
\begin{align}
    M^\text{sec}_\pm(t,\D E, \D E') = \delta_{\D E, \D E'} \F^\pm_{\D E, \T}(t).
\end{align}
The Liouville superoperator \eqref{LL} reduces then to the Lindblad form
which preserves positivity.
\begin{equation} \label{LL-sec}
\begin{split}
  &\t \L\, \rho = \\
  & \sum_{\substack{m,\al \\ \D E}} \left[ L_m^{\al}(\D E)\, \rho\, L_m^{\al}(\D E)\s 
  - \frac{1}{2} \{ L_m^{\al}(\D E)\s L_m^{\al}(\D E), \rho \} \right]
\end{split}
\end{equation}
with the secular Lindblad jump operators 
\begin{multline} \label{Lind-glob-sec}
  L^{\pm,\text{sec}}_{m}(t,\D E) = \sqrt{\Gamma_{m}} \times \\ \sum_{i,j} \d_{\pm \D E, E_i-E_j} \sqrt{\F^+_{E_i-E_j, \T}(t) }\, \ketbra{E_i}{E_i} c^\pm_{m} \ketbra{E_j}{E_j}
\end{multline}
defined for each energy difference $\D E = E_i - E_j$ appearing in the spectrum of the Hamiltonian $H_\text{S}$.
The double sum runs over all eigenstates $\ket{E_l}$ of the system Hamiltonian $H_\text{S}$ with eigenenergies $E_l$.
As we see, the secular approximation automatically removes the imaginary parts $G_{\Delta E,\T}(t)$.  

\subsection{Coherent approximation}

Because the secular approximation removes too much information and can miss important physics, 
we developed in \cite{B07-Dimer, B07-Trimer}, along the lines of \cite{Wichterich_coh,Kirsanskas+Wacker-CohLindPhenom, Davidovic2020, Nathan-CohLindDerivation}, the \textit{coherent approximation}   as the least invasive method of restoring positivity.
The arithmetic mean of the real parts in the off-diagonal terms of $\mathcal{M}^\pm_{\D E, \D E'}(t)$ is replaced by the geometric mean while the imaginary parts are removed%
\footnote{The replacement is motivated by the sign of the smallest eigenvalue of $\mathcal{M}^\pm_{\D E, \D E'}$ which is proportional to $\mathfrak{G}^2~-~\mathfrak{A}^2~-~\D G^2 \leq 0$ where $\mathfrak{G}$ is the geometric mean, $\mathfrak{A}$ is the arithmetic mean of $\F^\pm_{\D E, \T}(t)$ and $ \F^\pm_{\D E', \T}(t)$ (with $\mathfrak{G}^2 \leq \mathfrak{A}^2$) and $\D G = [{G_{\D E,\T}(t)- G_{\D E',\T}(t)}]/{2}$, present in its off--diagonals.
Replacing $\mathfrak{A}$ with $\mathfrak{G}$ and setting $\D G = 0$ in \eqref{Redfield-matrix} lifts the negative eigenvalue exactly to zero.}
\begin{multline}
       \mathcal{M}^{\pm,\text{coh}}_{\D E, \D E'}(t)  =\\ 
  \begin{pmatrix}
    \F^\pm_{\D E, \T}(t), & \sqrt{\F^\pm_{\D E, \T}(t)\F^\pm_{\D E', \T}(t)} \\
   \sqrt{\F^\pm_{\D E, \T}(t)\F^\pm_{\D E', \T}(t)}, & \F^\pm_{\D E', \T}(t)
  \end{pmatrix}.
\end{multline}
In this way, the negative eigenvalue gets shifted up to zero
while the diagonal elements, directly influencing energy populations, stay untouched.  
The Liouville operator \eqref{LL-sec} is then given in terms of the \textit{coherent} Lindblad jump operators
\begin{multline} \label{Lind-glob-coh}
  L^{\pm,\text{coh}}_{m}(t) = \\
  \sqrt{\Gamma_{m}} \sum_{i,j} \sqrt{\F^+_{E_i-E_j, \T}(t)}\, \ketbra{E_i}{E_i} c^\pm_{m} \ketbra{E_j}{E_j}.
\end{multline}
They are equal to coherent sums of the secular jump operators, $L^{\pm,\text{coh}}_{m}(t) = \sum_{\D E} L^{\pm,\text{sec}}_{m}(t,\D E)$ over the spectrum of the energy differences which motivates their name. 
For more details regarding their derivation and discussion of their properties we refer to \cite[App. A]{B07-Trimer}.

\section{Beyond the wideband limit} \label{app:nonwideband}

\begin{figure}[ht]
    \centering
    \includegraphics[width=0.9\linewidth]{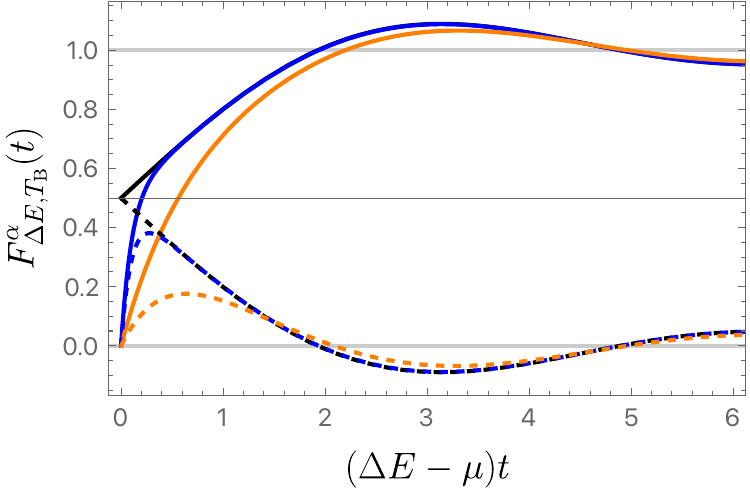}
    \caption{{Coefficients $F^\alpha_{\Delta E,\T}(t)$ with $\alpha=-$ (solid lines) and $\alpha=+$ (dashed lines) as functions of time. The black lines show $F^\alpha_{\Delta E,\T}(t)$ in the wideband limit (cf. \eqref{ReF}) while other lines show $F^\alpha_{\Delta E,\T}(t)$ for the Lorentzian \eqref{Loretz-Gamma} with $\Delta_\Gamma = 2(\Delta E-\mu) = 2/\tau_\text{c}$ (orange) or $\Delta_\Gamma = 10(\Delta E-\mu) = 10/\tau_\text{c}$ (blue). The temperature is $\T=0$ and $\Delta E-\mu>0$.}}
    \label{fig:non_wideband}
\end{figure}

{In the wideband limit, where $\Ga(\omega)$ is assumed constant everywhere, we find for the coupling coefficients the \textit{finite} short--time limit $\lim_{t\rightarrow t_0} A_m^\pm(t, \Delta E)= \Gamma_m/2$.
For any integrable function $\Ga(\omega)$, however, the coefficients must vanish at short times $\lim_{t\rightarrow t_0}A_m^\pm(t, \Delta E)=0$
since 
$|A_m^\pm(t,\Delta E)|~\leq ~\| \Ga_m \|_1 (t-t_0)/\pi $ with $\| \Gamma_m \|_1$ 
the $L^1$--norm. 
Consequently, the equations \eqref{ReA}--\eqref{ReF} and \eqref{ImA}--\eqref{ImF} cannot hold for very short times, $t-t_0\lesssim 1/\| \Ga_m \|_1$.
In the particular example of a Lorentz--like distribution
\begin{align} \label{Loretz-Gamma}
    \Gamma_m(\omega) & =\frac{\Gamma_m \Delta_\Gamma^2}{\Delta_\Gamma^2+(\Delta E-\omega)^2}, & \| \Ga_m \|_1 &= \pi\, \Ga_m \D_\Ga
\end{align} 
the width $\Delta_\Gamma$ determines the scale of agreement with the wideband limit, 
namely, for $t-t_0\!\gg\! 1/\D_{\Ga}$ the coefficients $A_m^\pm(t,\Delta E)$ become well approximated by the wideband limit (cf. Fig. \ref{fig:non_wideband}). 
{(For more general $\Gamma_m(\omega)$, the conditions hold analogously with $\D_{\Ga}$ determined by the variation of $\Gamma(\omega)$ \cite[Sec.~4]{LL-MTh}.)}
If additionally $\D_{\Ga} \gg 1/\tau_\text{c} = \max\left(\kB \T, |\D E - \mu|\right) \gtrsim \Ga_m$, the effects of the time--dependent temperature $T(t)$ become relevant.}

\section{Derivation of the $T(t)$ approximation} \label{app:TeffDeriv}

In order to derive the approximation \eqref{int-approx} we first need to introduce a useful similarity.

\subsection{Functional similarity} \label{app:Similarity}

\begin{figure}[t]
  \includegraphics[width=\linewidth]{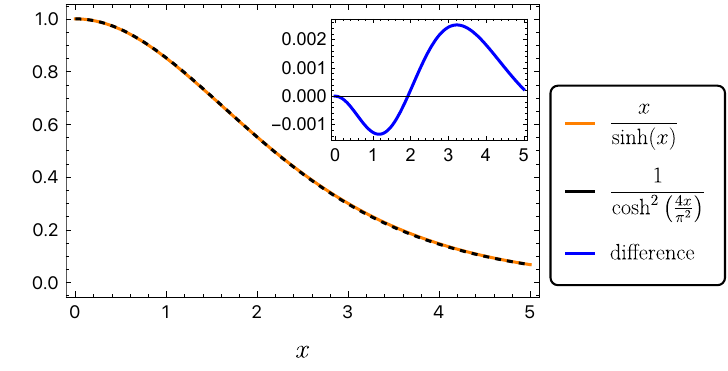}
  \caption{Astonishing similarity between $x/\sinh(x)$ (orange) and $1/\cosh^2(4x/\pi^2)$ (black) with their difference in the inset (blue).}
  \label{fig:Similarity}
  \includegraphics[width=\linewidth]{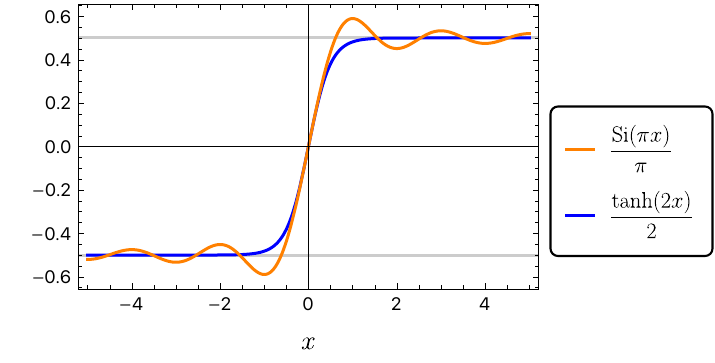}
  \caption{Comparison of ${\Si(\pi x)}/{\pi}$ (orange) with ${\tanh (2x)}/{2}$ (blue) as functions of $x$. The maximal difference amounts $\approx 0.11$ and vanishes for small and for large $x$.} 
  \label{fig:SinInt-Tanh}
  \includegraphics[width=\linewidth]{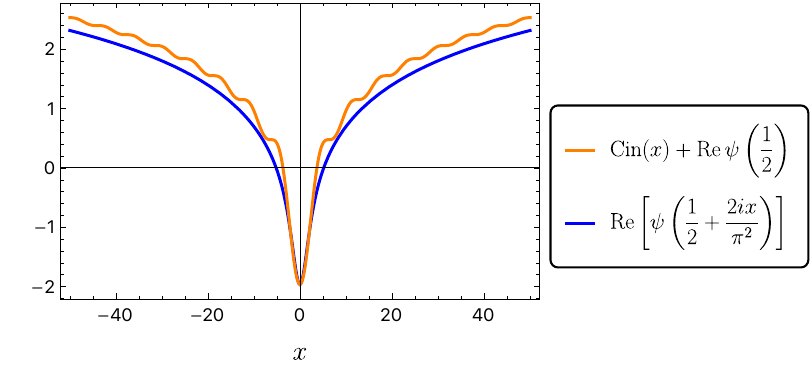}
  \caption{Comparison of  $\text{Cin}(x)+\text{Re}\left[\psi(1/2)\right]$ (orange) with  $\text{Re}\left[\psi\left(1/2+2i x/\pi^2\right)\right]$ (blue) as functions of $x$.} 
  \label{fig:Cin-Digamma}
\end{figure}

There holds an astonishing similarity between the two hyperbolic functions 
\begin{equation} \label{magic}
  \frac{x}{\sinh(x)} \approx \frac{1}{\cosh^2(4x/\pi^2)}, 
\end{equation}
as shown in Fig.~\ref{fig:Similarity}.
Despite intensive (re)search and discussion \cite{SE-Similarity} we did not succeed in providing any elementary proof of it%
\footnote{The Taylor expansions of the reciprocals of both functions are very close to each other and grow exponentially fast. By this, both functions must decay exponentially and their absolute difference quickly gets negligibly small. 
}. 
The constant $4/\pi^2$ is chosen by the requirement that both functions have equal integrals over $[0,\infty)$ {which will have a physical significance in our applications}.

\subsection{Estimation of the integral \eqref{int}}\label{app:TeffDeriv2}

\def\kB{}

\begin{figure}[t]
  \includegraphics[width=\linewidth]{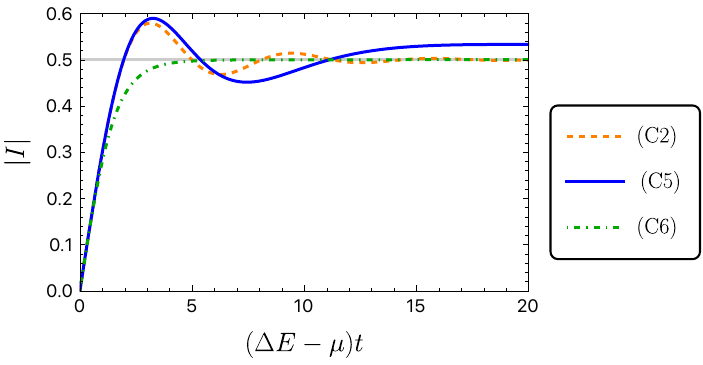}
  \caption{Compared are the exact value of the integral $I$ \eqref{defI} (orange), Si--tanh approximation \eqref{app_I_Si_tanh} (blue) and tanh--tanh approximation \eqref{I-approx-result} (green). The temperature of the bath is $\kB\T=0.08|\Delta E-\mu|$.
  } \label{fig:SinInt}
\end{figure}

Here, we will apply the discovered functional similarity \eqref{magic} to the integral \eqref{int}. 
First we split the integrand into a product and use the similarity \eqref{magic} to obtain 

\begin{align}
  I &= \kB \T \int_{0}^{t-t_0} \frac{\sin(\D E' \tau)}{ \sinh(\pi \kB \T\tau)} d\tau \label{defI} \\
  &= \frac{\D E'}{\pi} \int_{0}^{t-t_0} \frac{\pi \kB \T\tau}{\sinh(\pi \kB \T\tau)} 
    \frac{\sin(\D E' \tau)}{\D E' \tau} d\tau \nonumber \\
  &\approx \frac{\D E'}{\pi} \int_{0}^{t-t_0} \frac{1}{\cosh^2(4 \kB \T\tau/\pi)} 
    \frac{\sin(\D E' \tau)}{\D E' \tau} d\tau
\end{align}
with $\D E' \equiv \D E - \mu$ and omit the Boltzmann constant $\kB$ for shorter notation.
In order to calculate this integral analytically, we use in the second factor the following trick: 
We observe that the first factor contributes significantly only for $4 \kB \T\tau/\pi \lesssim 1$ (and is negligible otherwise)
where the $\tanh$--function is almost linear so that we can replace $\tau \approx \pi \tanh(4 \kB \T \tau/\pi)/(4 \kB \T)$.
This brings us to
\begin{equation}
  I \approx \frac{\D E'}{\pi} \int_{0}^{t-t_0} \frac{1}{\cosh^2\left(\frac{4 \kB \T\tau}{\pi}\right)}
    \frac{\sin\left(\frac{\pi \D E'}{4 \kB \T} \tanh\left(\frac{4 \kB \T \tau}{\pi}\right)\right)}{\frac{\pi \D E'}{4 \kB \T} \tanh\left(\frac{4 \kB \T \tau}{\pi}\right)} d\tau
\end{equation}
which, by substituting $u = \frac{\pi \D E'}{4 \kB \T} \tanh\left(\frac{4 \kB \T \tau}{\pi}\right)$, leads to the result 
\begin{equation}  \label{app_I_Si_tanh}
  I \approx \frac{1}{\pi} \Si\left[\frac{\pi \D E'}{4 \T} \tanh\left( \frac{4 \T (t-t_0)}{\pi} \right) \right],
\end{equation}
with $\Si$ being the sine integral function. 
The result inherits, however, the disadvantage of the exact coefficient $F_{\D E, \T}(t)$ which overshoots beyond the allowed range, namely, $|I|$ increases above the value $1/2$ which leads to problems (cf. Sec.~\ref{sec:TimeDepCoef}). 
But its behavior at early and late times $t$ as well as small and large $\D E$ values is the same as in the tanh-formula \eqref{int-approx} which is free of that problem. 
Treating the excess oscillations (cf. Fig.~\ref{fig:SinInt}) as an error of the first order approximation (in the coupling between the system and the bath) we want to correct it to physically acceptable range bearing physical interpretation. 
For this sake we observe that the sine integral (Si) function is close to 
$\Si(\pi x)/\pi \approx \tanh(2 x)/2$ which builds the Fermi function and stays in the proper range (cf. Fig.~\ref{fig:SinInt-Tanh}). 
That replacement delivers the final approximation
\begin{equation} \label{I-approx-result}
  I \approx \frac{1}{2} \tanh\left[ \frac{\D E'}{2 \kB \T} \tanh \left(\frac{4 \kB \T (t-t_0)}{\pi} \right) \right]
\end{equation}
which, technically, presents our main result \eqref{int-approx}. 
The connection to the Fermi function is an essential point which allows us to interpret the final result, used in $F^\pm_{\D E, \T}(t)$ \eqref{ReF}, as the Fermi distribution with modified, time--dependent temperature $T(t)$ \eqref{Teff}. 

An alternative way to motivate this approximation \cite[Ch. 3]{EK-PhD} 
{(much simpler to derive but not uniform in $\D E$)} 
is to introduce the effective temperature
by matching the slopes of the Fermi distribution and of the coefficient {$F^-_{\D E, \T}(t)$} at $\D E = \mu$ 
\begin{equation}
\begin{split}
  \frac{1}{4 \kB T(t)} &= \left. \frac{\p F^-_{\D E, \T}(t)}{\p \D E} \right|_{\D E=\mu} 
  = \int_0^{t-t_0} \frac{\kB \T \tau\; d\tau}{\sinh(\pi \kB \T \tau)}.
  \end{split}
\end{equation}
Using again the \textit{astonishing similarity} $x/\sinh(x) \approx 1/\cosh^2(4x/\pi^2)$ (cf.~\eqref{magic} in  App.~\ref{app:Similarity} ), (\ref{Teff}) follows. 

\def\kB{k_\text{B}}

\subsection{Estimation of the integral \eqref{ImF}} \label{app:TeffDeriv3}
\def\kB{}

Applying the same approximations as discussed in Sec. \ref{app:TeffDeriv2} to the integral \eqref{ImF}, we derive  
\begin{align}
  & G_{\Delta E, \T}(t)=\kB \T \int_{0}^{t-t_0} \frac{1-\cos(\D E' \tau)}{ \sinh(\pi \kB \T\tau)} d\tau \nonumber\\
  &\approx \frac{\D E'}{\pi} \int_{0}^{t-t_0} \frac{\pi \kB \T\tau}{\cosh^2(4 \kB \T\tau/\pi)} 
    \frac{1-\cos(\D E' \tau)}{\D E' \tau} d\tau \nonumber\\
    &\approx \frac{\D E'}{\pi} \int_{0}^{t-t_0} \frac{1}{\cosh^2\left(\frac{4 \kB \T\tau}{\pi}\right)}
    \frac{1-\cos\left(\frac{\pi \D E'}{4 \kB \T} \tanh\left(\frac{4 \kB \T \tau}{\pi}\right)\right)}{\frac{\pi \D E'}{4 \kB \T} \tanh\left(\frac{4 \kB \T \tau}{\pi}\right)} d\tau\nonumber\\
    &=\frac{1}{\pi} \text{Cin}\left[\frac{\pi \D E'}{4 \T} \tanh\left( \frac{4 \T (t-t_0)}{\pi} \right) \right]\label{appGDeltaapprox}
\end{align}
with $\D E' \equiv \D E - \mu$ and omitting the Boltzmann constant $\kB$ for shorter notation. The cosine integral function is defined by
\begin{equation}
    \text{Cin}(x)=\int_0^x \frac{1-\cos(t)}{t}\text{d}t
\end{equation}
and is close to $\text{Cin}(x)\approx \text{Re}\left[\psi\left(1/2+2i x/\pi^2\right)-\psi(1/2)\right]$ (cf. Fig. \ref{fig:Cin-Digamma}) where $\psi$ is the digamma function \cite[Sec. 6.3]{abramowitz1968handbook} which builds $G_{\Delta E,\T}(\infty)$ \cite[Ch. 3]{EK-PhD}. Using this replacement, we end up with the approximation
\begin{multline}
        G_{\Delta E,\T}(t)\approx \\ \frac{1}{\pi}\text{Re}\Biggl[\psi\left(\frac{1}{2}+i \frac{\D E'}{2 \pi \kB \T} \tanh \left(\frac{4 \kB \T (t-t_0)}{\pi} \right) \right)-\psi\left(\frac{1}{2}\right)\Biggr]
\end{multline}
which is identical to \eqref{ImF_eff_T}.

\def\kB{k_\text{B}}


\section{Correlations between system and bath} \label{app:exactCorrelationswithbath}

\begin{figure}[t]
\includegraphics[width=\linewidth]{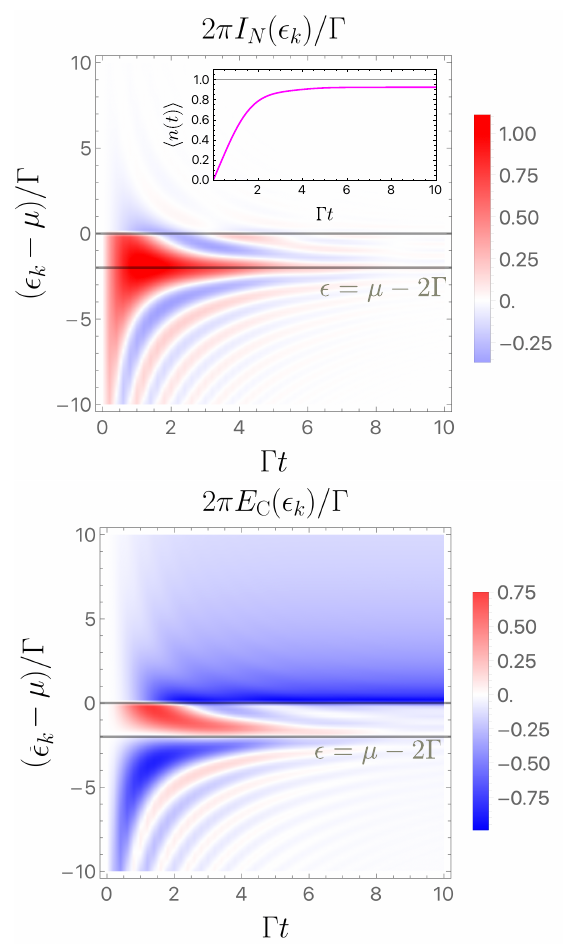}
  \caption{Particle current $I(\eps_{k})$ (top) and the coupling energy $E_\text{C}(\eps_{k})$ (bottom) as functions of the energy of the bath modes, $\eps_k$, and time, $t$.
  The inset (top) shows the particle number of the system as a function of time. 
  We consider a system given by the Hamiltonian \eqref{HU=0} with the parameters $\T=0$ and $\eps=\mu-2\Gamma < \mu$ and an initially empty dot. The oscillations of off--resonant tunneling processes are given by $e^{i(\eps_k-\eps)t}$.}
  \label{fig:IN+EC}
\end{figure}
\begin{figure}[t]
\includegraphics[width=\linewidth]{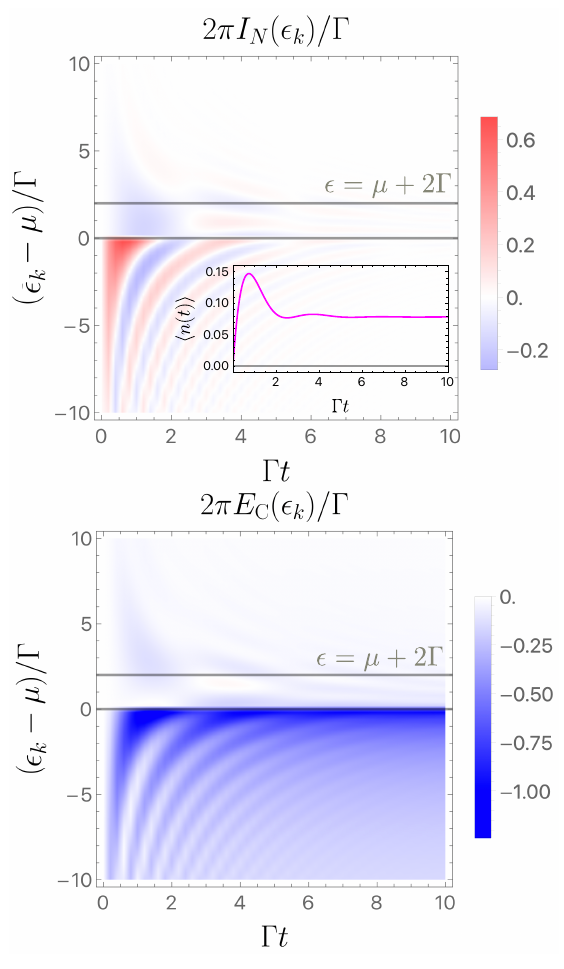}
  \caption{Same as in Fig.~\ref{fig:IN+EC} with the energy $\eps=\mu+2\Gamma > \mu$.}
  \label{fig:IN+EC_egmu}
\end{figure}

For the non--Coulomb interacting system described by \eqref{HU=0}, it is possible to give not only an exact expression of the quantum dot annihilation operator, $c_\text{H}(t)$ (cf. \eqref{cexact}), but also an exact expression of the bath annihilation operators
\begin{equation}
    b_{k,\text{H}}(t)=e^{-i \eps_{k} t} b_{k} -i\, \overline{\gamma_{k}} \int_0^t c_{\text{H}}(s)e^{-i \eps_{k}(t-s)}\text{d}s, 
    \label{app_aexact}
\end{equation}
derived from the Heisenberg's equation of motion.
This enables us to calculate the correlation $\left<c^\dagger b_k\right>$ between the quantum dot and a bath mode with energy $\eps_k$,
of which the imaginary part gives the single--mode particle current
\begin{equation}
    I_N(\eps_{k})=2\,\mathrm{Im}\left[\gamma_k\left<c^\dagger b_{k}\right>\right]
\label{I(eq)}
\end{equation}
from the bath mode $k$ to the quantum dot. 
Summing $I_N(\eps_k)$ over all modes $k$ in the bath gives the total particle current satisfying the balance equation, $\partial_t\left<c^\dagger c\right>=\sum_{k} I(\eps_k)$. 
The real part of the correlation corresponds to the single--mode coupling energy 
\begin{equation}
    E_\text{C}(\eps_{k})=2\,\mathrm{Re}\left[\gamma_k\left<c^\dagger b_{k}\right>\right]
    \label{EC(eq)}
\end{equation}
and the sum over all modes $k$ in the bath gives the total coupling energy, $\left<H_\text{C}\right>=\sum_{k} E_\text{C}(\eps_{k})$.

Let us first consider an initial state different from the ground state of the quantum dot at $\T=0$ (in contrast to Fig. \ref{fig: exactvsRedfield}) in order to better see the relaxation process to the ground state. 
Let $\eps < \mu $ and an initial particle number be $\left<n(0)\right>=0$. 
In Fig. \ref{fig:IN+EC}, the quantities $I_N(\eps_k)$ and $E_\text{C}(\eps_{k})$ are shown as functions of time and energy, $\eps_k$. 
From the conservation of energy for the quantum dot and bath alone, $H_\text{S} + H_\text{B}$, the current would flow only at the energy level $\eps_k=\eps$ but by including also the coupling energy, $H_\text{C}$,
the off--resonant currents at $\eps_k\neq\eps$ may appear.
Bath electrons with $\eps_k<\eps$ reduce the coupling energy $E_\text{C}(\eps_{k})$ (cf. blue shading in Fig. \ref{fig:IN+EC}, bottom) while electrons with $\eps_k>\eps$ increase the coupling energy $E_\text{C}(\eps_{k})$ (cf. red shading in Fig. \ref{fig:IN+EC}, bottom) in time. 
These processes lead to relaxation towards the ground state of the quantum dot, $\left<n\right>=1$ (cf. Fig. \ref{fig:IN+EC}, inset), with a small deviation, $\left<n\right><1$, caused by the off--resonant tunneling processes from the quantum dot back to bath states above the Fermi edge at $\eps_k = \mu$. 
In order to satisfy the energy conservation, the coupling energies $E_\text{C}(\eps_{k})$ for $\eps_k > \mu$ must become negative (cf. blue shading in Fig. \ref{fig:IN+EC}, bottom).  
In consequence, the final state lies energetically higher than the ground state. 

Finally, we consider a situation similar to the one shown in Fig.~\ref{fig: exactvsRedfield}
with the quantum dot initially in its ground state $\left<n(0)\right>=0$ for $\eps>\mu$.
Here, the tunneling processes are not peaked around $\eps_k\approx\eps$ but around the Fermi level with off--resonant current from states below the Fermi level into the quantum dot (cf. red shading in Fig.~\ref{fig:IN+EC_egmu}, top). This also leads to an energy increase in the quantum dot that is compensated by the decrease of the coupling energies $E_\text{C}(\eps_k)$ to negative values for states below the Fermi edge (cf. blue shading in Fig. \ref{fig:IN+EC_egmu}, bottom). 

\FloatBarrier

\bibliography{qdots}

\end{document}